%% file: PaperDraft.tex
\documentclass[sigconf]{acmart}
\AtBeginDocument{%
  \providecommand\BibTeX{{%
    \normalfont B\kern-0.5em{\scshape i\kern-0.25em b}\kern-0.8em\TeX}}}
    
\usepackage{subcaption}
\usepackage{multirow}
\usepackage{algorithm}
\usepackage{algorithmic}
\usepackage{booktabs}
\usepackage{caption}

\usepackage{amsmath}

\makeatletter
\def\@ACM@copyright@check@cc{}
\makeatother
\setcopyright{cc}
\setcctype{by}
\copyrightyear{2025}
\acmYear{2025}
\acmConference[CHI '25]{CHI Conference on Human Factors in Computing Systems}{April 26-May 1, 2025}{Yokohama, Japan}
\acmBooktitle{CHI Conference on Human Factors in Computing Systems (CHI '25), April 26-May 1, 2025, Yokohama, Japan}\acmDOI{10.1145/3706598.3714249}
\acmISBN{979-8-4007-1394-1/25/04}

\begin{document}

\title[AlphaPIG]{AlphaPIG: The Nicest Way to Prolong Interactive Gestures in Extended Reality}

\author{Yi Li}
 \affiliation{%
   \institution{Monash University}
   \city{Melbourne}
   \country{Australia}
 }
 \email{yi.li5@monash.edu}

  \author{Florian Fischer}
 \affiliation{%
   \institution{University of Cambridge}
   \city{Cambridge}
   \country{United Kingdom}
 }
\email{fjf33@cam.ac.uk}

  \author{Tim Dwyer}
 \affiliation{%
   \institution{Monash University}
   \city{Melbourne}
   \country{Australia}
 }
\email{Tim.Dwyer@monash.edu}

  \author{Barrett Ens}
 \affiliation{%
   \institution{University of British Columbia}
   \city{British Columbia}
   \country{Canada}
 }
 \email{barrett.ens@ubc.ca}
 
 \author{Robert Crowther}
 \affiliation{%
   \institution{University of New England}
   \city{Armidale}
   \country{Australia}
 }
\email{rcrowth2@une.edu.au}

  \author{Per Ola Kristensson}
 \affiliation{%
   \institution{University of Cambridge}
   \city{Cambridge}
   \country{United Kingdom}
 }
\email{pok21@cam.ac.uk}

  \author{Benjamin Tag}
 \affiliation{%
   \institution{University of New South Wales}
   \city{Sydney}
   \country{Australia}
 }
 \email{benjamin.tag@unsw.edu.au}
 
\renewcommand{\shortauthors}{Li et al.}
\begin{abstract}

%In the current paper, we propose a toolkit, AlphaPIG, to address this concern in future interaction design. AlphaPIG takes advantage of the latest shoulder fatigue model, NICER, to guide the future XR prototype with a pair of parameters: alpha and beta. Alpha will enable adaptive fatigue manipulation in real-time interaction, which avoids the tedious parameter tuning for a particular application. Beta, at the same time, will control the timing of activating the manipulation without interrupting the main task. In the case study of using AlphaPIG and the Go-Go technique in a mid-air selection task, results show that the condition with Go-Go+AlphaPIG received lower physical fatigue than the condition of using the Go-Go technique only. Meanwhile, participants perceived a similar level of ownership between the control and treatment conditions. We believe that AlphaPIG can assist future XR development by making prolonged gestural interaction and an engaging user experience.

Mid-air gestures serve as a common interaction modality across Extended Reality (XR) applications, enhancing engagement and ownership through intuitive body movements. However, prolonged arm movements induce shoulder fatigue—known as "Gorilla Arm Syndrome"—degrading user experience and reducing interaction duration. Although existing ergonomic techniques derived from Fitts' law (such as reducing target distance, increasing target width, and modifying control-display gain) provide some fatigue mitigation, their implementation in XR applications remains challenging due to the complex balance between user engagement and physical exertion. We present \textit{AlphaPIG}, a meta-technique designed to \textbf{P}rolong \textbf{I}nteractive \textbf{G}estures by leveraging real-time fatigue predictions. AlphaPIG assists designers in extending and improving XR interactions by enabling automated fatigue-based interventions. Through adjustment of intervention timing and intensity decay rate, designers can explore and control the trade-off between fatigue reduction and potential effects such as decreased body ownership. We validated AlphaPIG's effectiveness through a study (N=22) implementing the widely-used Go-Go technique. Results demonstrated that AlphaPIG significantly reduces shoulder fatigue compared to non-adaptive Go-Go, while maintaining comparable perceived body ownership and agency. Based on these findings, we discuss positive and negative perceptions of the intervention. By integrating real-time fatigue prediction with adaptive intervention mechanisms, AlphaPIG constitutes a critical first step towards creating fatigue-aware applications in XR.

\end{abstract}

\begin{CCSXML}
<ccs2012>
   <concept>
       <concept_id>10010147.10010178.10010179.10003352</concept_id>
       <concept_desc>Computing methodologies~Information extraction</concept_desc>
       <concept_significance>300</concept_significance>
       </concept>
   <concept>
       <concept_id>10003120.10003121.10003122</concept_id>
       <concept_desc>Human-centered computing~HCI design and evaluation methods</concept_desc>
       <concept_significance>500</concept_significance>
       </concept>
   <concept>
       <concept_id>10003120.10003121.10003122.10003334</concept_id>
       <concept_desc>Human-centered computing~User studies</concept_desc>
       <concept_significance>300</concept_significance>
       </concept>
 </ccs2012>
\end{CCSXML}

\ccsdesc[300]{Computing methodologies~Information extraction}
\ccsdesc[500]{Human-centered computing~HCI design and evaluation methods}
\ccsdesc[300]{Human-centered computing~User studies}

\keywords{Mid-air Interaction, Adaptive Interaction, Gorilla Arm, Shoulder Fatigue}


\maketitle

\input{sections/1_introduction}

\input{sections/2_background}
\input{sections/3_design}

\input{sections/4_casestudy}
\input{sections/5_discussion}
\input{sections/6_conclusion}
\input{sections/7_openscience}

% \begin{acks}
% LLM 
% \end{acks}

\bibliographystyle{ACM-Reference-Format}
\bibliography{Reference}

%\appendix
%\input{appendix/1_part_one}

%\input{appendix/1_interview}
% \input{appendix/2_part two}
% \input{appendix/3_online resources}

\end{document}

%% file: sections/1_introduction.tex
\section{Introduction}

Extended reality (XR)---which encompasses virtual (VR) and augmented reality (AR)---depends on mid-air gestures involving the whole torso, arms, and hands to provide a full and natural interaction experience. These upper-body movements involve various muscles across the hand, forearm, upper arm, shoulder, and torso. Prolonged activity can quickly tire these muscles and lead to significant physical fatigue, particularly in the shoulder joint, a condition often referred to as “Gorilla Arm Syndrome”~\cite{palmeira2023quantifying}. This, in consequence, negatively affects the user experience, interaction time, and overall engagement, posing a critical challenge in designing XR interfaces~\cite{laviola20173d}. Despite a variety of techniques for improving ergonomics being introduced, e.g., applying a linear or non-linear gain to input devices translation, rotation, or scaling~\cite{dai2024portal,liu2022distant,wentzel2020improving}, their implementation in XR remains limited. An ongoing challenge is the lack of guidance for practical implementation of these interaction techniques, coupled with a critical need for rigorous assessment of their effectiveness in reducing physical strain and potential interaction-related side effects~\cite{maslych2024research}, such as diminished body ownership, compromised task performance, and reduced user engagement.

Meanwhile, modelling shoulder fatigue in mid-air interactions has emerged as an important research domain in recent years. Since the initial formulation of the Consumed Endurance (CE) model in 2014~\cite{Hincapie-Ramos_CE_2014}, researchers have developed four successive iterations~\cite{jang2017modeling, villanueva2023advanced, Li_revisitingCE2023}, culminating in the most recent NICER model proposed in 2024~\cite{li2024nicer}.

These models can predict user fatigue in real-time, offering significant potential for adaptive systems that dynamically improve interaction ergonomics. This approach is similar to the adaptive design in exertion games, where game content (e.g., object movement speed, task frequency, or attack force) is adjusted dynamically based on body exertion~\cite{nenonen2007using, she2020hiitcopter,montoya2020enhancing}. However, previous techniques depend on adjusting the behaviour of game objects, which limits their use in XR applications where intrinsic content is primarily static (e.g., training simulations of physical environments or immersive visualisation applications). What is needed is an \textit{application- and task-independent} approach for adapting user fatigue in real-time interactions.

Building on these advances in fatigue modelling and adaptive exertion games design, we developed \textit{AlphaPIG}, a meta-technique designed to \textbf{P}rolong \textbf{I}nteractive \textbf{G}estures and guide designers toward creating fatigue-aware XR designs. At its core, AlphaPIG provides a mechanism that adaptively modulates the parameters and properties of a chosen interaction technique; we thus refer to it as a \textit{meta-technique}. AlphaPIG offers easy integration of adaptive, real-time fatigue management without task interruption. The two meta-parameters contribute to generating a single scalar, $\alpha$, which dynamically adjusts interaction based on predicted user fatigue. AlphaPIG provides designers with a platform to implement, test, and refine various techniques that subtly intervene in user behaviour by integrating real-time fatigue predictions directly into the XR environment. Moreover, unlike previous methods, AlphaPIG is task-independent and enables dynamic adjustment of user-interaction techniques. This novel approach provides a unique opportunity to systematically explore trade-offs between physical fatigue and other critical dimensions of user experience, ultimately facilitating a more comprehensive optimization of interactive systems. 

We demonstrate the applicability of AlphaPIG using the ubiquitous target pointing task with the well-established Go-Go interaction technique.\footnote{Go-Go is a VR interaction method allowing users to interact with distant objects by extending their virtual hand beyond the physical hand reach~\cite{poupyrev1996go}.}
We compare physical fatigue, body ownership, and control in the following three conditions: (1) baseline direct interaction; (2) Go-Go interaction with default parameters recommended by the authors~\cite{poupyrev1996go}; and (3) AlphaPIG-assisted Go-Go with nine combinations of different intervention timing and intensity. Our findings from $22$ participants show how the proposed meta-technique can be used to explore a trade-off between mitigating physical fatigue and managing potential interaction side effects, lowering the perception of body ownership in the observed case. Moreover, our approach successfully identified an improved implementation of Go-Go that significantly reduces fatigue while preserving comparable body ownership. These results highlight the potential of AlphaPIG to create personalized, fatigue-aware user experiences and validate its capability as a critical, ready-to-use tool for future XR applications.

The key contributions of this work are as follows:

 \begin{description}
    \item[Meta-Technique] We introduce \textit{AlphaPIG}, a \emph{meta-technique} that modifies existing interaction techniques, allowing designers to manage user fatigue effectively.  
    \item[Tool] We present the AlphaPIG Unity Plugin (``AlphaPIG API''), a ready-to-use tool designed for XR developers, enabling seamless plug-and-play integration of the AlphaPIG meta-technique to manage fatigue and adapt interactivity in XR environments.
    \item[Case] We demonstrate the applicability, efficacy, and utility of AlphaPIG through the implementation and evaluation of redirected pointing enhanced with fatigue awareness. Our case study provides ecological validity by showcasing AlphaPIG’s superior performance in reducing physical fatigue while preserving user ownership compared to the widely used Go-Go manipulation technique.
\end{description}

Integrating real-time fatigue management in XR interfaces marks a transformative step towards more user-centric design practices in the XR field. By empowering designers to create interactions that account for physical strain, AlphaPIG can improve the longevity and overall enjoyment of XR applications, making them more viable for long-term use. This work paves the way for future research on adaptive, fatigue-aware interactions, contributing to a broader understanding of ergonomics in digital environments and enhancing the user experience across diverse XR applications, such as gaming, virtual workspaces, and rehabilitation.

%% file: sections/2_background.tex
\section{Related Work}
This research was inspired by exertion games designed to enhance user engagement through real-time adaptation of exertion levels. Additionally, it draws from studies that developed innovative interaction techniques to minimize shoulder fatigue in mid-air interactions. 

\subsection{Exertion Adaptation in Exertion Games Design}
\label{sec:exergame}

Designing exertion-adapted interactions can effectively increase user engagement in exertion games (exergames) -- gameplay experiences that involve intensive body movement \cite{karaosmanoglu2024born}. Typically, this is achieved through adaptive, customised manipulation of a set of game properties, including object moving speed~\cite{nenonen2007using}, over-head target placement~\cite{she2020hiitcopter}, task frequency~\cite{masuko2006fitness}, or attack force~\cite{montoya2020enhancing}, as soon as physiological measures exceed a specific exertion threshold. For instance, \citet{nenonen2007using} used real-time heart rate to adjust the skiing speed during a shooting exergame, while HIITCopter~\cite{she2020hiitcopter} integrated analogue heart rate signals into VR high-intensity interval training. Similarly, \citet{masuko2006fitness} defined three gameplay intensities based on players' maximum heart rates as a pre-game adaptation method, and \citet{montoya2020enhancing} developed a VR upper-body exergame that adapts muscle contraction via surface electromyography (sEMG) signals to encourage exercise toward a target exertion level. Although these applications demonstrate the effectiveness of physiological feedback in dynamically controlling game environments~\cite{munoz2019kinematically}, they remain task-dependent, limiting their applicability to broader, non-gaming mid-air interactions in static environments. Our proposed meta-technique is a generalizable approach with no constraints in tasks and applications.

\subsection{Shoulder Fatigue Manipulation in Mid-air Interaction}
\label{sec:novelmanipulation}

\begin{figure}
    \centering
    \includegraphics[width=\linewidth]{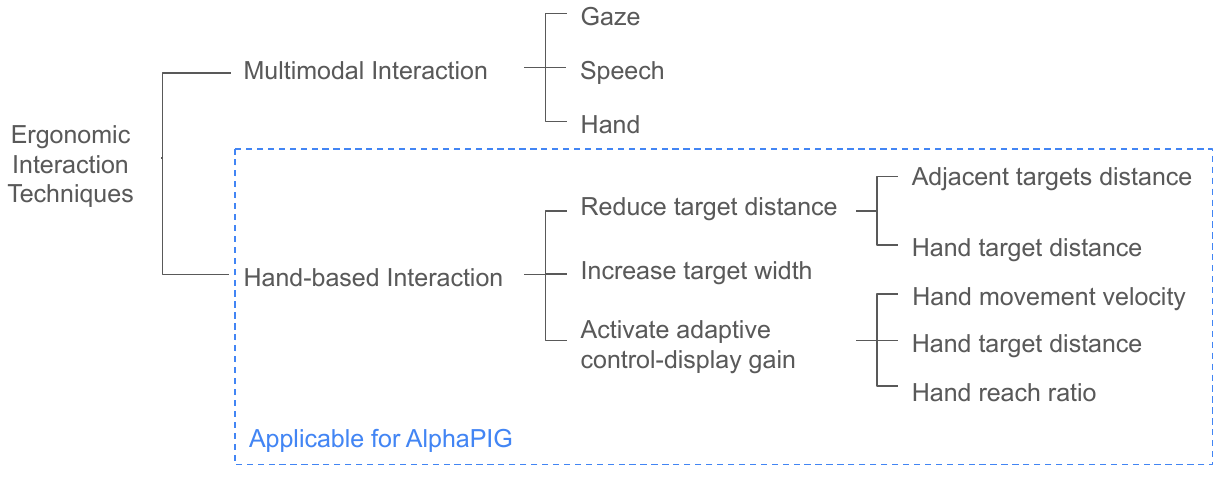}
    \caption{Classification of physically ergonomic interaction techniques in XR by modalities and manipulation strategies. Methods within the blue dashed box are suitable applications of AlphaPIG.}
    \Description[Classification of physically ergonomic XR interaction methods.]{Classification of physically ergonomic XR interaction methods. This hierarchical diagram categorizes ergonomic interaction techniques into two main branches: multimodal and hand-based interactions. Multimodal interaction encompasses gaze, speech, and hand modalities. Hand-based interaction includes three strategies: reducing target distance (subdivided into adjacent targets distance and hand target distance), increasing target width, and activating adaptive control-display gain (comprising hand movement velocity, hand target distance, and hand reach ratio). AlphaPIG applications are specifically noted as applicable to the hand-based interaction techniques.}
    \label{fig:interactionmethodtaxonomy}
\end{figure}

Significant efforts have been dedicated to developing novel interaction techniques that minimize the need for large, extensive body movements to prolong mid-air gestural interaction and enhance user engagement. These ergonomic methods can be separated into two categories based on the number of user input channels (see Figure \ref{fig:interactionmethodtaxonomy}). The first category consists of multimodal techniques, which use several alternative input modalities, such as gaze or speech, to  avoid the need for precise mid-air gestures. In a study evaluating the performance of multiple interaction modalities in target selection and object manipulation tasks, \citet{wang2021interaction} found that combining gaze, speech, and gesture resulted in a significantly lower physical workload than gesture-only interactions. While the present unimodal approach focuses exclusively on hand-based interaction, we anticipate that our contributions and methodology will provide valuable foundations and insights for future multimodal interaction research.

The second category of ergonomic methods focuses on unimodal hand-based interaction techniques, where input is generated exclusively through hand movements or handheld devices. Fatigue is mitigated through three key strategies guided by the principles of Fitt's law~\cite{balakrishnan2004beating} to (1) reduce the target distance, (2) increase
the target width, and (3) activate adaptive control-display gain to reduce extended arm movements.

\noindent\textbf{(1) Reduce target distance}.
Strategies to minimize the distance between the end effector and the target can be divided into two main approaches. The first approach is to reduce the distance between neighbouring targets. Empty areas in the target geometry can be minimized~\cite{dachselt2006survey} or skipped by early end-point prediction~\cite{yu2019modeling}. The second approach is to decrease hand target distance. ProxyHand~\cite{iqbal2021reducing} was proposed to reduce the distance between the physical hand and the virtual targets by introducing a 3D spatial offset to the hand representation. The XRgonomics toolkit~\cite{Belo_XRgonomics_2021} optimizes 3D UI placement based on the user comfort score derived from averaging CE~\cite{Hincapie-Ramos_CE_2014}, muscle activation~\cite{bachynskyi2015informing}, and Rapid Upper Limb Assessment (RULA)~\cite{mcatamney1993rula}. This toolkit allows users to select the interaction areas and recommends the most comfortable position relative to the body. One significant limitation of this system is its inability to respond to real-time fatigue scores, and it only offers guidance for target placement. To address this gap, our work proposes the development of a toolkit that incorporates real-time fatigue measurements to guide general interaction tasks more effectively.

\noindent\textbf{(2) Increase target width}. Alternatively, high interaction efficiency and reduced fatigue can be achieved by increasing the size of the target object or the selector tool~\cite{poupyrev1998egocentric}. Examples in 2D interactions are Flashlight~\cite{liang1994jdcad} or Aperture Selection~\cite{forsberg1996aperture}, which expand the coverage area of the selection tool to indirectly reduce the gap between adjacent targets. Improvement of task selection time was also found in expanding 3D targets in VR~\cite{yu2018target,shi2023expanding}.

\noindent\textbf{(3) Adaptive control-display gain}.
Additionally, immersive 3D environments provide opportunities to influence user behaviour by dynamically adjusting the control-display (CD) gain. The CD gain is the ratio between the movement of input devices (control space) and the movement of controlled targets (display space)~\cite{balakrishnan2004beating}. In isomorphic 3D interactions where the CD gain equals 1, the movement maintains a one-to-one relationship between control and display spaces when unaffected by external manipulation. An anisomorphic CD gain (CD gain $\neq 1$) alters the relationship between the user's control space and display space, directly affecting body movements. When 0 $<$ CD gain $<$ 1, the translation and rotations of hand movements are amplified before being applied to the selected target. thereby reducing the arm movements required for interaction. Erg-O~\cite{montano2017erg} is a manipulation technique that enables virtual objects to remain in the same position within the display space while being accessible in the control space. By employing reduced control-display (CD) gain, it ensures the reachability of distant objects in virtual space with minimal physical effort. Similarly, Ownershift~\cite{feuchtner2018ownershift} expands the control space by redirecting the virtual hand, allowing users to maintain overhead interactions while keeping their physical hand in a less fatiguing position.

Fatigue reduction can be achieved by transitioning from isomorphic to anisomorphic CD gains in ergonomic interaction techniques, with transitions triggered by hand movement velocity, hand target distance, or hand reach ratio. For instance, Relative Pointing~\cite{vogel2005distant} applied a velocity-based CD gain of 0.7, allowing for large movements with subtle hand movement while preserving high selection precision. Building on this concept, an anisomorphic raycasting technique~\cite{andujar2007anisomorphic} dynamically adjusts the CD gain based on proximity to the target, increasing the gain when the input device is far from the target object to reduce interaction time. The Go-Go technique~\cite{poupyrev1996go} takes a different approach by modifying the CD gain based on the user’s elbow extension, allowing users to interact with objects beyond their physical reach. As the arm extends, the offset between physical and virtual hands increases proportionally, reducing physical exertion by requiring less elbow extension.

The novel interaction techniques above significantly reduce physical workload compared to default gestural direct manipulation. Yet, they have been only sparsely implemented in practice. One major reason for this is their potential negative impact on other aspects of user experience~\cite{guo2023interpretative}. For example, techniques that introduce a distance offset between the virtual and physical hands, such as those described in~\cite{feuchtner2018ownershift,poupyrev1996go,dai2024portal}, can decrease body ownership, leading to phenomena like the rubber hand illusion~\cite{botvinick1998rubber}. As a first step to address these challenges and maintain a sense of control during such manipulations, \citet{wentzel2020improving} employed a Hermite spline to quantify the trade-off between the amplified manipulation and the body ownership, revealing a decreasing trend in body ownership as manipulation intensity increases. However, the trade-off between fatigue reduction and other aspects of user experience remains unexplored. Moreover, to the best of our knowledge, there are currently no tools or techniques available to facilitate such investigations.

This paper presents the first practical tool that can assist the exploration of these trade-offs between physical fatigue and other dimensions of user experience by modifying the manipulation timing and amplification, ultimately maximising the usability and applicability of ergonomic interaction techniques in future interface designs.

%% file: sections/3_design.tex
\section{AlphaPIG: A Meta-Technique and Tool for Creating Fatigue-Aware XR Applications}
\label{sec:design}

To automatically adapt XR interactions to user's fatigue, we introduce \textbf{\textit{AlphaPIG}, a novel meta-technique that dynamically adjusts selected interaction techniques based on real-time fatigue predictions}. 

The core idea of AlphaPIG is based on the observation that fatigue can be systematically controlled through ergonomic hand-based interaction methods using scalars across three intervention levels. At the primary level, scalars modulate manipulation intensity by strategically adjusting interaction parameters to amplify the translation, rotation, and scaling of input devices~\cite{liu2022distant} (e.g., modifying the position or size of an ``input zone'' used for gesture recognition~\cite{Kristensson12}). The secondary level focuses on manipulation timing, determining the precise moment of intervention. For example, Erg-O ~\cite{montano2017erg} defined a threshold for retargeting the virtual hand position. Lastly, at the tertiary level, scalars control manipulation intensity and timing simultaneously. For instance, in the Go-Go technique~\cite{poupyrev1996go}, the intervention is dynamically defined by the spatial relationship between hand and shoulder. Virtual hand remapping is triggered when the hand's reach exceeds two-thirds of the arm length, with intervention intensity linearly proportional to the distance surpassing this threshold. While interaction parameters in literature have proven effective in improving task performance, the chosen values are typically based on simple heuristics. For example, interaction parameters may be based on a predicted user's physical comfort zone, or they may be ``optimized" for a specific interaction task. Such fixed parameters, which make assumptions about tasks, reduce generalisability and practical applicability.

To address these limitations, we propose dynamically adjusting a selected manipulation variable $\theta$
of the interaction technique based on the user’s physiological state. This adaptive method draws inspiration from the Alpha Blending technique in image processing~\cite{smith1995alpha}, dynamically adjusting interaction parameters in response to real-time physiological feedback. We achieve this by leveraging a robust and validated fatigue model that predicts user fatigue over time using motion tracking, eliminating the need for direct physiological sensing methods, such as EMG or EEG~\cite{villanueva2023advanced, li2024nicer, jang2017modeling}.
Notably, our AlphaPIG meta-technique enables researchers and interaction designers to define a direct relationship between user fatigue and the input method, independent of the considered interaction task or XR application. It essentially creates a closed feedback loop between the system and the user's physical state: as fatigue accumulates over time, the interaction automatically adjusts to alleviate the increased fatigue. It thus offers a general solution for intervening in any movement-based XR application, improving compatibility with a wide range of interaction techniques and paradigms. 

To make AlphaPIG accessible to the community, we developed the \textbf{AlphaPIG Unity API} to assist in creating fatigue-aware XR applications. This API allows XR designers to explore and test various meta-parameters of AlphaPIG, enabling them to dynamically adjust and optimize interaction parameters for their custom applications. In the following sections, we provide a detailed description of the computation and implementation of AlphaPIG.

\subsection{Comprehensive Fatigue Model: NICER}
While AlphaPIG is not limited to a specific fatigue model, we use NICER---the state-of-the-art, validated and publicly available model---for demonstrating and evaluating its capabilities~\cite{li2024nicer}. We use the public Unity API of NICER in AlphaPIG to guide the management of real-time shoulder fatigue. NICER, which stands for \textit{New and Improved Consumed Endurance and Recovery Metric of Mid-Air Interactions}, is a validated model that predicts real-time shoulder fatigue based on dynamic arm movements~\cite{li2024nicer}. It employs a hybrid biomechanical muscle contraction and shoulder torque model, distinguishing between interactions with different hand travel paths and interaction durations. Building on the previous NICE model~\cite{Li_revisitingCE2023}, NICER introduces a recovery factor, enhancing its effectiveness compared to other established fatigue models (e.g., CE~\cite{Hincapie-Ramos_CE_2014}, CF~\cite{jang2017modeling,villanueva2023advanced}) and aligning more closely with subjective ground truth, as demonstrated in a comparison of tasks with different exertion levels~\cite{li2024nicer}. 

\subsection{Dynamic Intervention Workflow}
\label{sec:intervention-workflow}

\begin{figure}
    \centering
    \includegraphics[width=\linewidth]{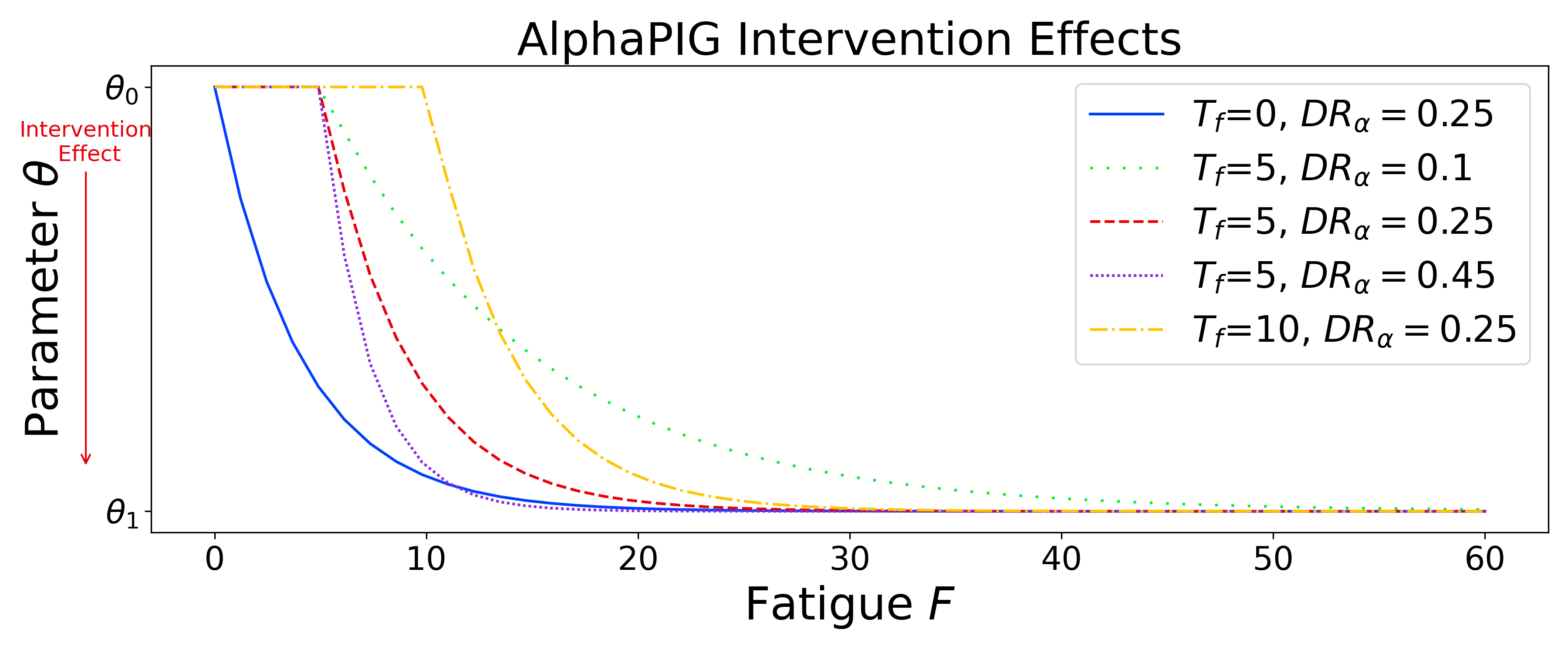}
    \caption{AlphaPIG provides two meta-parameters, $T_f$ and $DR_\alpha$, for testing different fatigue intervention effects. A higher fatigue threshold $T_f$ raises the fatigue level at which the interaction technique parameter $\theta$ is adjusted. The decay rate $DR_\alpha$ determines to what level a small increase in fatigue (above the threshold $T_f$) affects $\theta$. A small decay rate (e.g., $DR_\alpha$ = 0.1, green dotted line) results in a larger intervention effect for the same fatigue level $F$ than a medium ($DR_\alpha$ = 0.25, red dashed line) or large decay rate (such as $DR_\alpha$ = 0.45, purple loosely dotted line). The horizontal shift between lines indicates a delay in the intervention effect until the user exceeds the target fatigue: $T_f$=0 (blue solid line) triggers intervention immediately, $T_f$=5 (red dashed line) delays intervention until a higher fatigue level is reached, and $T_f$=10 (yellow dash-dotted line) further delays intervention until the user is even more fatigued.}
    \Description[Comparison of fatigue intervention effects using different meta-parameter combinations.]{Comparison of fatigue intervention effects using different meta-parameter combinations. This graph illustrates how parameter theta changes across fatigue levels (0-60) under varying conditions. Five curves demonstrate different combinations of fatigue threshold (Tf) and decay rate (DRa): immediate intervention (Tf=0, DRa=0.25) shows earliest response, while delayed interventions (Tf=5 and Tf=10) activate at progressively higher fatigue levels. The decay rate DRa influences intervention intensity, with lower rates (0.1) producing more gradual parameter adjustments compared to higher rates (0.45). All curves eventually converge at high fatigue levels, indicating similar long-term adaptation regardless of initial intervention timing.}
    \label{fig:alphaPIG-intervention-mechanism}
\end{figure}

We use an exponential decay function to determine how a slight increase in fatigue affects the interaction technique parameter. This approach provides relatively large interventions immediately after exceeding the threshold, ensuring timely adjustments. The first parameter of AlphaPIG, $DR_\alpha$, represents the \textit{decay rate} used to change the shape of the exponential decay function.
As can be inferred from Figure~\ref{fig:alphaPIG-intervention-mechanism}, a low decay rate (such as $DR_\alpha$ = 0.1, green dotted line) results in a smaller intervention effect for the same fatigue level $F$ than a medium ($DR_\alpha$ = 0.25, red dashed line) or large decay rate (such as $DR_\alpha$ = 0.45, purple loosely dotted line). The second parameter of AlphaPIG, $T_f$, is the target fatigue level that allows delay of the intervention effect until the user is "exhausted enough"; see the horizontal shift between $T_f$=0 (blue solid line), $T_f$=5 (red dashed line), and $T_f$=10 (yellow dash-dotted line).

After designers or researchers select one (or more) XR interaction technique(s) $I$, specifically for their potential to mitigate shoulder fatigue and precisely define one manipulation variable $\theta$ associated with each selected technique $I$, the exponential decay function will be used to customise the interaction across the three previously described intervention levels.

To ensure efficient exploration of the range of $\theta$ and maximize fatigue mitigation through significant intervention effects, designers or researchers should provide values of $\theta_{0}$ and $\theta_{1}$ to define the upper and lower bounds of the manipulation variable. The value $\theta_{0}$ corresponds to the default value used without intervention. The value $\theta_{1}$ is set to a value required for maximal intervention when user fatigue is at its expected highest. In the example of applying hand redirection to mitigate shoulder fatigue, $\theta$ represents the CD gain. A reasonable choice of $\theta_{0}$ will be 1, which ensures a direct 1:1 mapping between the physical and virtual hand movements if the interaction technique allows it. Conversely, a small value of $\theta_{1}$ corresponds to a scenario where even minimal hand movements near the body are substantially amplified. In the following, we assume that \textbf{lower $\theta$ values produce stronger intervention effects, resulting in greater fatigue reduction}.

Once the manipulation variable $\theta$ and its boundaries $\theta_{0}$ and $\theta_{1}$ have been chosen, AlphaPIG will dynamically intervene $\theta$ during real-time interaction using the scalar $\alpha$, according to the following equations:%
\begin{align}
    \theta_t &= \theta_{0} + \alpha_t (\theta_{1} - \theta_{0}),
    \label{eq:intervention-workflow-1}\\
    \text{with} \quad \alpha_{t} &= \begin{cases} 
    1 - e^{-DR_{\alpha} (F_{t} - T_{f})} &\text{if} \quad F_{t} \geq T_{f}, \\
    0 &\text{otherwise.}
    \end{cases}
    \label{eq:intervention-workflow-2}
\end{align}

Here, $\theta_t$ corresponds to the value the interaction technique parameter $\theta$ takes at time $t$ and only depends on a single scalar $\alpha_t$ that changes over time. If $\alpha_t = 0$, the interaction technique parameter takes its default value (i.e., $\theta_t=\theta_{0}$); for $\alpha_t = 1$, the boundary value $\theta_{1}$ is assumed (i.e., $\theta_t=\theta_{1}$). By setting $\alpha_t$ to values between $0$ and $1$, i.e., linearly interpolating between $\theta_0$ and $\theta_1$, parameter values between $\theta_{0}$ and $\theta_{1}$ can be achieved.

Eq.~\eqref{eq:intervention-workflow-2} describes how $\alpha_t$ changes over time, depending on the user's fatigue.
Here, $F_t$ denotes the user fatigue predicted by NICER at time $t$, while $T_f$ denotes the constant fatigue threshold chosen by the designer or researcher. 
When the user's fatigue is below the threshold $T_f$, $\alpha_t = 0$ and consequently $\theta_t = \theta_{0}$ holds, no intervention being implemented. 
As soon as the user's fatigue exceeds the threshold (i.e., $F_t \geq T_f$), an intervention is triggered, dynamically adjusting the interaction parameter $\theta_t$ to a value interpolated between the upper bound $\theta_{0}$ and the lower bound $\theta_{1}$ (see Figure~\ref{fig:alphaPIG-intervention-mechanism}). In general, progressively higher fatigue levels $F_t$ exceeding the threshold trigger increasingly substantial intervention effects. Crucially, the interventions occur at fairly small time intervals (utilizing a default frame rate of 60 Hz), enabling the parameter $\theta_t$ to be continuously and dynamically adjusted in real-time as long as the fatigue remains above the predefined threshold $T_f$.

In summary, the following meta-parameters define the AlphaPIG intervention workflow:
\begin{itemize}
    \item fatigue threshold $T_f$: the fatigue level at which the intervention starts,
    \item decay rate $DR_{\alpha}$: the extent to which a small change in fatigue (beyond $T_f$) affects $\theta$.
    \item manipulation variable $\theta_t$: the interaction technique parameter to be modified (e.g., the spatial offset between the control and the display space) at time $t$,
    \item manipulation bounds $\theta_{0}$, $\theta_{1}$: the default value ("no intervention") and the most extreme value ("maximum intervention") of the manipulation variable $\theta$,
\end{itemize}

$\theta$ represents the manipulation variable adjusted during the AlphaPIG intervention workflow, with its value bounded by $\theta_0$ (no intervention) and $\theta_1$ (maximal intervention). $\alpha$, expressed through the decay rate $DR_{\alpha}$, dictates how sensitively $\theta$ responds to fatigue levels exceeding the threshold $T_f$.

\subsection{Additional Consideration for Smooth Intervention Onset}
\label{sec:consideration_continuous}

The transition of control-display ratio between isomorphic (CD gain $= 1$) and anisomorphic (CD gain $\neq 1$) needs to be carefully managed to maintain interaction fluidity. For example, when the real-time fatigue level exceeds the fatigue threshold (i.e. $F_{t} \geq T_{f}$), implementing hand redirection without appropriate smoothing mechanisms results in discontinuous virtual hand movement, where the virtual hand position suddenly diverges significantly from the physical input position in the control space. To prevent this perceptual disruption and preserve the user's sense of embodiment and control, AlphaPIG implements an additional smoothening factor $\beta$ that enables gradual interpolation between the two statuses for a transition period.

\begin{equation}
    \label{eq:beta-smoothening}
    % p_v^{t} =  p_r^{t} + \beta_t * (F(p_r) - p_r).
    p_v^{t} =  p_v^{(\theta_0)} + \beta_t \cdot \left(p_v^{(\theta_t)} - p_v^{(\theta_0)}\right)
    % p_v =  I_{\theta_0}(p_r) + \beta_t * (I_{\theta_{t}}(p_r) - I_{\theta_0}(p_r)).
\end{equation}

Given the tracked position of the input device in the control space $p_r$ and starting at $\beta_t$=0, %\footnote{In practice, $p_v$ may also encapsulate non-positional information on the virtual user state, e.g., avatar orientation, or tool size.} 
the position in display space $p_v$ is gradually shifted from the baseline position $p_v^{(\theta_0)}=I_{\theta_0}(p_r)$ (isomorphic) to the virtual position determined by the manipulated interaction technique $p_v^{(\theta_t)}=I_{\theta_t}(p_r)$ (anisomorphic).

We have chosen to increase $\beta_t$ with a step size of $0.005$ once the fatigue threshold is exceeded, as this results in a transition period of 3.33 seconds for the constant 60Hz frame rate used, which is approximately the average time between two consecutive eye blinks~\cite{zenner2021blink}.

% Starting at $\beta_t$=0, \textit{once the fatigue threshold is exceeded} we gradually increase $\beta_t$ at each time step, using a step size of $0.005$. Since we use a constant 60Hz frame rate, this results in a linear increase (and thus an initial delay of intervention effects) over 3.33 seconds, roughly the average time between two consecutive eye blinks~\cite{zenner2021blink}. In other words, at the beginning of the intervention, the effects are delayed by up to 3-4 seconds.

\subsection{AlphaPIG API}

The AlphaPIG Unity API is a module interface that enables designers or researchers to integrate the dynamic intervention workflow described above into their XR applications with chosen fatigue measurements and desired interaction techniques. At its core is the \texttt{AlphaPIG} script, which triggers and updates interventions in a selected interaction technique based on real-time fatigue predictions. The API gives access to the scalar $\alpha$ after end users define the meta parameters, including the manipulation bounds $\theta_0$ and $\theta_1$ (``Default Value'' and ``Max Intervention Value''), the fatigue threshold $T_f$, and the decay rate $DR_{\alpha}$ in the graphical inspector. The API includes two callback functions, \texttt{SetInteractionTechnique} and \texttt{SetFatigueModel}, allowing custom input devices and other physiological fatigue measures to be passed to the adaptive system, improving the flexibility and practicability of AlphaPIG. %This means designers or researchers need to define the manipulation variable $\theta$ 

%% file: sections/4_casestudy.tex
\section{Case: Mid-air Selection Tasks}
\label{sec:casestudy}

In the following case, we demonstrate how AlphaPIG guides the integration of hand redirection into the ubiquitous mid-air pointing task in XR. We compare 11 conditions across three interaction techniques: the default technique, the Go-Go technique~\cite{poupyrev1996go}, and AlphaPIG-assisted Go-Go with nine parameter combinations.

\subsection{Interaction Technique: Go-Go}
\label{sec:redirection-technique}

We selected the well-established Go-Go technique~\cite{poupyrev1996go} as the interaction technique to be manipulated for mitigating shoulder fatigue during mid-air interaction tasks. The Go-Go technique is one of the pioneering approaches in hand space transformation, and it is considered ergonomic, as remapping physical hand positions $p_r^{t}$ at time $t$ in Eq.~\eqref{eq:go-go} reduces elbow extension during arm movement~\cite{mcatamney1993rula}.

\begin{equation}\label{eq:go-go}
    % \resizebox{0.91\hsize}{!}{
    I(p_r^{t}, \theta_t) = \begin{cases}\left(\Vert p_r^{t} \Vert_2 + k \cdot (\Vert p_r^{t} \Vert_2 - D)^2\right) \cdot \left(\frac{p_r^{t}}{\Vert p_r^{t} \Vert_2}\right) &\text{\footnotesize if $~ \Vert p_r^{t} \Vert_2 \geq D,$} \\ &\text{\footnotesize where $D=\theta_t \cdot L$,}  \\
    p_r^{t} &\text{\footnotesize otherwise.}
    \end{cases}
    % }
\end{equation}

Here, $\Vert p_r^{t} \Vert_2$ denotes the distance between the hand and the shoulder at time $t$; in particular, Go-Go introduces offsets only in the direction the arm is pointing (i.e., it constitutes an isotropic technique). Go-Go relies on two parameters: $D$ and $k$, defining the distance between hand and shoulder at which the redirection is onset and the size of the redirection effect, respectively. When the hand reach is within the threshold, i.e., $\Vert p_r^{t} \Vert_2 < D$, the virtual hand position maintains a 1:1 mapping with the physical hand. Otherwise, the virtual hand will be shifted away from the physical hand’s position according to a non-linear (quadratic) mapping. The reference point was shifted from the chest to the shoulder, following the protocol outlined in \cite{chan2022investigating}. 

To simplify the demonstration in the paper, we only consider manipulating the distance threshold $D$ while fixing $k = \frac{1}{12}$ (as suggested in the original study~\cite{poupyrev1996go}). As such, the manipulation parameter $\theta$ at time $t$ modifies the distance threshold $D$ relative to the participant's arm length $L$: $D=\theta_t \cdot L$ (with $\theta=\frac{2}{3}$ being the original value used in the Go-Go implementation described in~\cite{poupyrev1996go}). The parameter bounds are set to $\theta_{0}=1$ and $\theta_{1}=\frac{1}{6}$, respectively. 

\begin{figure}
    \centering
    \includegraphics[width=\linewidth]{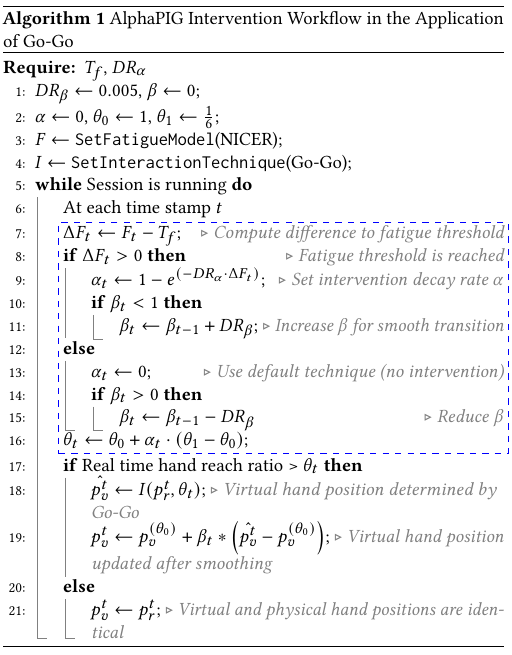}
    %\caption{AlphaPIG Intervention Workflow in the Application of Go-Go}
    \Description[]{}
    \label{alg:AlphaPIG}
\end{figure}

Algorithm 1 depicts the implementation of the AlphaPIG intervention workflow in Unity with the chosen Go-Go technique. The code within the blue dashed box (lines 7 to 16) corresponds to Eq.~\eqref{eq:intervention-workflow-1} and~\eqref{eq:intervention-workflow-2}. 
At each time step $t$, it updates both the fatigue $F_t$ and the manipulation parameter $\theta_t$. The interaction technique with modified parameter $\theta_t$ is then used to update the pre-smoothing position of the input device in display space ($\hat{p_v^{t}}$) with Eq~\eqref{eq:go-go}. In the end, users perceive the post-smoothing virtual hand position $p_v^{t}$ updated using the baseline position in the control space by Eq~\eqref{eq:beta-smoothening}.

The manipulation variable $\theta$ critically determines when the virtual hand begins to diverge from the physical hand and affects how much the arm must be extended to reach the distant target position. Existing literature~\cite{wentzel2020improving,rietzler2018breaking} has identified side effects of this beyond-the-real hand extension on body ownership. In AlphaPIG-assisted Go-Go, a low $T_f$ will trigger an early starting time of such extension, and a low $DR_{\alpha}$ will induce a subtle spatial offset between physical and virtual hand positions. Consequently, AlphaPIG can intervene in manipulation timing and intensity—representing the most complex intervention level described in Section~\ref{sec:design}. Moreover, \textbf{we aim to demonstrate that using fatigue to modulate manipulation intensity with AlphaPIG can transform the prior trade-off between \textit{redirection offset} and \textit{body ownership} into a new trade-off between \textit{physical fatigue} and \textit{body ownership}, while simultaneously identifying an ``optimal`` approach to balancing these factors effectively.}

\subsection{User Study}
We conducted this case study to understand how the 2D design space of parameters $T_{f}$ and $DR_{\alpha}$ influence the intervention in $\theta$ according to Eq.~\eqref{eq:intervention-workflow-1} and Eq.~\eqref{eq:intervention-workflow-2} by comparing them with the \textit{Default} interaction ($\theta=1$) and the standard Go-Go technique ($\theta=\frac{2}{3}$). 

We include three levels of the fatigue threshold $T_f$ to systematically investigate the \textit{Timing} effect of introducing AlphaPIG-assisted Go-Go at three distinct stages of the interaction task. We define these levels relative to the participants’ maximum fatigue value, obtained from the \textit{Default} interaction during a training task, to account for individual variations in fatigue. For example, if a participant experiences a fatigue level of 20\% at the end of the \textit{Default} interaction task, $F_{\text{max}} = 20\%$ will be used as user-specific reference value. We used the following levels: $T_f=0$ (\textit{start}), $T_f=25\% \cdot F_{\text{max}}$ (\textit{quarter}) and $T_f=50\% \cdot F_{\text{max}}$  (\textit{mid}) to approximate interventions at the start, the first quarter, and the middle of the interaction, respectively, in the study conditions. 

The second parameter $DR_{\alpha}$ controls the \textit{Decay Speed} of the manipulation parameter $\theta$ (i.e., the Go-Go hand reach ratio) upon reaching the fatigue threshold  $T_f$. We evaluated three levels of $DR_{\alpha}$: 0.1, 0.25, and 0.45, which we respectively denoted as \textit{low}, \textit{medium}, and \textit{high} in our study conditions. 

Based on the relationship between parameters and intervention effects visualized in Figure~\ref{fig:alphaPIG-intervention-mechanism}, and the known trade-off between intervention effects and ownership, we hypothesize the following study outcomes:

\begin{description}
    \item[H1] There is a positive relationship between \textit{Timing} and user cumulative fatigue (i.e., later \textit{Timing} results in higher fatigue).
    \item[H2] There is a negative relationship between \textit{Decay Speed} and user cumulative fatigue (i.e., higher \textit{Decay Speed} results in lower fatigue).
    \item[H3] There is a positive relationship between \textit{Timing} and body ownership (i.e., later \textit{Timing} results in higher ownership).
    \item[H4] There is a negative relationship between \textit{Decay Speed} and body ownership (i.e., higher \textit{Decay Speed} results in lower ownership).
\end{description}

In addition, we hope to identify at least one combination of parameters \textit{Timing} and \textit{Decay Speed} in AlphaPIG that improves over the original Go-Go technique in terms of fatigue while preserving a similar ownership perception, offering an optimal solution to address this trade-off.

\paragraph{Dependent Measures}
The nine combinations of different $T_f$ and $DR_{\alpha}$ levels ($3 \times 3 = 9$) allow us to explore various intervention intensities and timings. To assess their impact on physical fatigue and body ownership, we measure both real-time shoulder fatigue using the NICER model and \textit{Task Completion Time (TCT)} in seconds. To obtain a robust estimate of final physical fatigue accumulated for each trial, we calculated the mean of all fatigue predictions from the last second and noted ``Cumulative Fatigue'' in subsequent analyses. 

After each condition, participants completed a five-item questionnaire to self-report their perceptions of body ownership and control, which included the following questions:
\begin{description}
    \item[Q1] "There were times when I felt that the virtual hand was part of my body." 
    \item[Q2] "There were times when I felt like I had more than one right hand."
    \item[Q3] "There were times when I felt I could control the virtual hand as if it was my own." 
    \item[Q4] "I never felt that the virtual hand was part of my body." 
    \item[Q5] "Do you notice any changes during the interaction time?"
\end{description}

Questions 1 to 4 are adapted from the Ownershift study~\cite{feuchtner2018ownershift} and use a 7-point Likert-type scale from $-3$ to $3$, where $-3$ corresponded to "Strongly Disagree" and $3$ to "Strongly Agree". Question 5 is open-ended to explore whether participants noticed the intervention and how it impacted their perception.

\paragraph{Participants}
Following approval from our institution's Human Ethics Research Committee, we recruited 22 participants (12 male, 10 female) with a mean age of 27.5 years (sd: 8.4). Only right-handed participants were included to ensure a consistent study setup. Recruitment was conducted through word of mouth. Of the 22 participants, 20 had prior experience with VR interactions, making most participants familiar with 3D target selection tasks and the sensation of full body ownership in VR.

\paragraph{Apparatus}
We used a Meta Quest 3 HMD, powered by a computer running on an Intel Core i7-7920X CPU and equipped with an NVIDIA GeForce GTX 1080. The upper body tracking was supported by the Meta Movement SDK\footnote{https://developer.oculus.com/documentation/unity/move-overview/}.

\paragraph{Procedure and Task}
After each participant completed the online consent form and the demographic survey, we measured their arm's length to calibrate the virtual hand and position targets accordingly. The task required selecting 40 targets randomly distributed across a $9 \times 9$ curved grid, with each increment set at 7.5\textdegree. The grid featured inclination angles ranging from -30\textdegree~to 30\textdegree ~relative to the participant's chest, azimuth angles from 60\textdegree~to 120\textdegree~relative to the participant's chest, and radial distance of one unit of the participant's arm length. Participants were instructed to stand throughout the entire study session to ensure body stability. Each trial began when the participant virtually pressed a green button in front of their chest. Participants were then required to use their right index fingertips to select individual target buttons. Upon correct selection, the target button disappeared, and the next target was activated. All targets were activated in a fixed order to enable direct comparison between conditions. A training session consisting of one trial using the \textit{Default} interaction technique was conducted to measure participants’ maximum fatigue scores ($F_{\text{max}}$) during the mid-air selection task. Each trial of 40 targets lasted approximately one minute, and a two-minute break was given between trials, during which participants completed the post-study survey. The entire study session, including the training session and 11 study conditions with order determined by a balanced Latin Square for a within-subject design, took approximately 40 minutes.

\subsection{Study Results}
% Our study analysis had two objectives: (1) to verify the trade-off between fatigue and body ownership, and (2) to determine whether AlphaPIG could identify Go-Go parameters that significantly reduce fatigue while preserving body ownership. 

Our study analysis had two objectives: (1) to examine the effects of AlphaPIG's meta-parameters—intervention timing and intensity decay speed—on cumulative fatigue, task completion time (TCT), and sense of body ownership and agency; and (2) to evaluate AlphaPIG's impact on interaction experience compared to baseline direct interaction and non-adaptive Go-Go techniques. For the first objective, we conducted a non-parametric, two-way ANOVA using Aligned Rank Transformation (ART)~\cite{Wobbrock11} to test the main and interaction effects of \textit{Timing} $\times$ \textit{Decay Speed}. Post-hoc comparisons were conducted using the ART-C algorithm~\cite{elkin2021ARTC} with Holm-Bonferroni correction to identify pairwise significant differences. This non-parametric approach was necessary as Shapiro-Wilk tests indicated non-normal distribution for all measures (cumulative fatigue, TCT, questionnaire responses; all $p$<$0.001$). For the second objective, we performed a non-parametric one-way analysis using \textit{Interaction Technique} as a single factor with 11 groups, employing Friedman tests to identify differences between \textit{Default}, \textit{Go-Go}, and the nine AlphaPIG conditions. For post-hoc analyses, exact probability values were computed using the Wilcoxon distribution due to the small sample size, with Holm-Bonferroni correction applied to control for multiple comparisons. The adjusted p-values are reported at conventional significance levels (e.g., 0.05, 0.01, 0.001).

\paragraph{\textbf{Cumulative Fatigue}}

\begin{figure}[!h]
    \centering
    \includegraphics[width=0.49\linewidth]{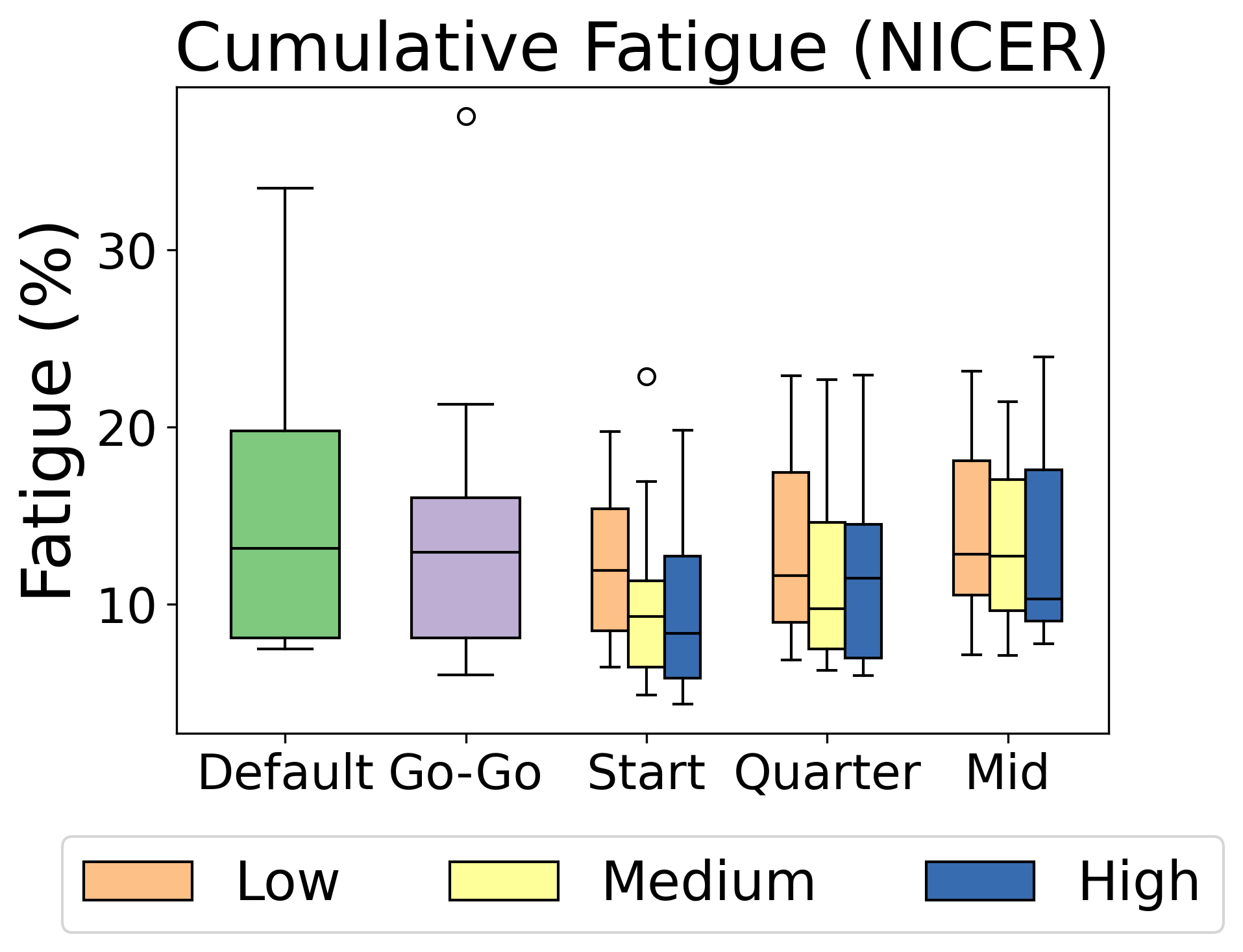}
    \includegraphics[width=0.49\linewidth]{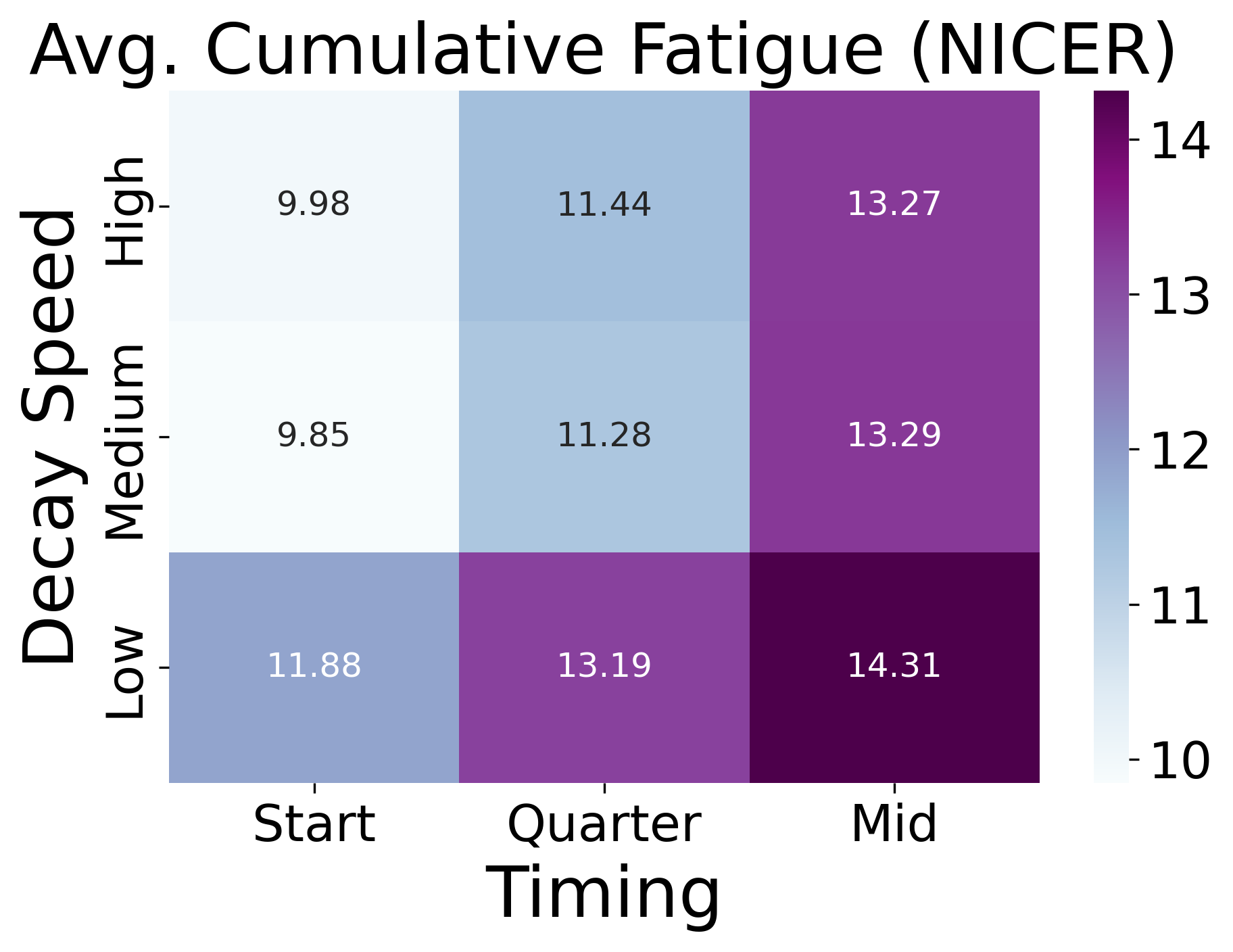}
    \caption{\textbf{Left:} Cumulative fatigue values grouped by \textit{Timing} (x-axis) and \textit{Decay Speed} (color-coded), with baseline conditions \textit{Default} and \textit{Go-Go} included for comparison (shown in green and purple boxes, respectively).
    \textbf{Right:} Mean cumulative fatigue values across all participants, plotted by \textit{Timing} (x-axis) and \textit{Decay Speed} (y-axis).}
    \Description[Comparison of cumulative fatigue across timing conditions and decay speeds.]{Comparison of cumulative fatigue across timing conditions and decay speeds. Two complementary visualizations present NICER fatigue data: a box plot showing distribution of fatigue values and a heatmap displaying mean fatigue levels. The box plot demonstrates fatigue percentages ranging from approximately 5-35\%, with baseline conditions (Default and Go-Go) showing greater variability than experimental conditions (Start, Quarter, Mid). The heatmap reveals an interaction between timing and decay speed, with lowest average fatigue (9.85\%) occurring at medium decay speed during start timing, and highest fatigue (14.31\%) at low decay speed during mid timing. Fatigue generally increases from start to mid timing conditions across all decay speeds, with low decay speed consistently producing higher fatigue values.}
    \label{fig:fatigue_comparison}
\end{figure}

Significant main effects were found for both \textit{Timing} ($F_{2,168}=55.41$, p $<$ 0.001) and \textit{Decay Speed} ($F_{2,168}=21.79$, p $<$ 0.001) on predicted cumulative fatigue. No significant interaction effect was observed between these meta-parameters ($F_{4,168}=0.48$, p = 0.75). Post-hoc tests revealed that fatigue increased significantly with \textit{Timing}, with the lowest levels observed in the "Start" condition and highest in the "Mid" condition. Significant differences in fatigue were found between "Low" and "Medium" decay speeds, as well as between "Low" and "High" decay speeds (statistics are given in Table~\ref{tab:fatigue_stats_ART_ANOVA_posthoc}). Although no significant difference was found between "Medium" and "High" decay speeds, cumulative fatigue generally decreased as \textit{Decay Speed} increased. Figure~\ref{fig:fatigue_comparison} presents both the distribution (left) and mean values (right) of predicted cumulative fatigue across all nine AlphaPIG conditions, as well as baseline conditions \textit{Default} and \textit{Go-Go} for comparison.

A significant effect of \textit{Interaction Technique} on cumulative fatigue was found ($\chi^2_{10}=92$, p $<$ 0.001, N $=$ 242). Two AlphaPIG conditions demonstrated significantly lower fatigue compared to \textit{Go-Go}: \textit{Start-Medium} (p $<$ 0.01) and \textit{Start-High} (p $<$ 0.01). Additionally, three conditions exhibited significantly lower fatigue than \textit{Default}: \textit{Start-Medium} (p $<$ 0.01), \textit{Start-High} (p $<$ 0.05) and \textit{Quarter-Medium} (p $<$ 0.01). No AlphaPIG conditions showed significantly higher fatigue than either baseline condition, and no significant difference in cumulative fatigue was found between \textit{Default} and \textit{Go-Go} (p = 1).\footnote{An overview of all post-hoc test results is given in Table~1 in the Supplementary Material.}

% Post-hoc analyses revealed that \textit{Start-Medium} demonstrated significantly lower fatigue compared to all conditions of "Mid" intervention timing (\textit{Mid-Low}: p $<$ 0.01; \textit{Mid-Medium}: p $<$ 0.01; \textit{Mid-High}: p $<$ 0.001) and \textit{Quarter-Low} (p $<$ 0.001). Similar results were observed for the \textit{Start-High} condition (\textit{Mid-Low}: p $<$ 0.01; \textit{Mid-Medium}: p $<$ 0.01; \textit{Mid-High}: p $<$ 0.01; \textit{Quarter-Low}: p $<$ 0.01) and the \textit{Quarter-Medium} condition (\textit{Mid-Low}: p $<$ 0.01; \textit{Mid-Medium}: p $<$ 0.05; \textit{Mid-High}: p $<$ 0.01; \textit{Quarter-Low}: p = 0.088). Additionally, \textit{Start-Medium} exhibited significantly lower fatigue than \textit{Start-Low} (p $<$ 0.05).

These results demonstrate that AlphaPIG significantly reduces cumulative fatigue compared to \textit{Default} and the non-adaptive \textit{Go-Go} through systematic adjustment of two meta-parameters: \textit{Timing} and \textit{Decay Speed}.

\begin{figure}
    \centering
    \includegraphics[width=0.49\linewidth]{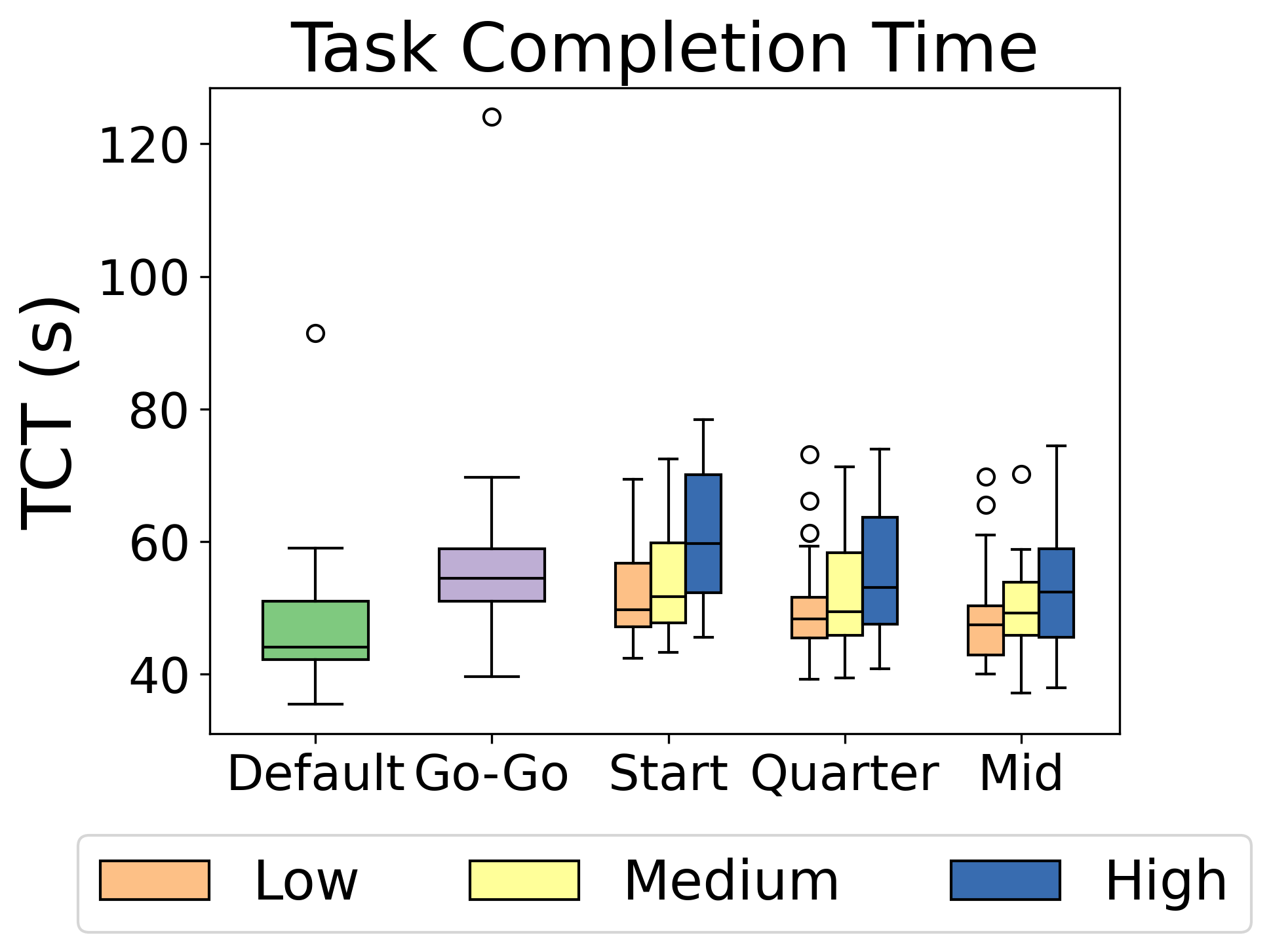}
    \includegraphics[width=0.49\linewidth]{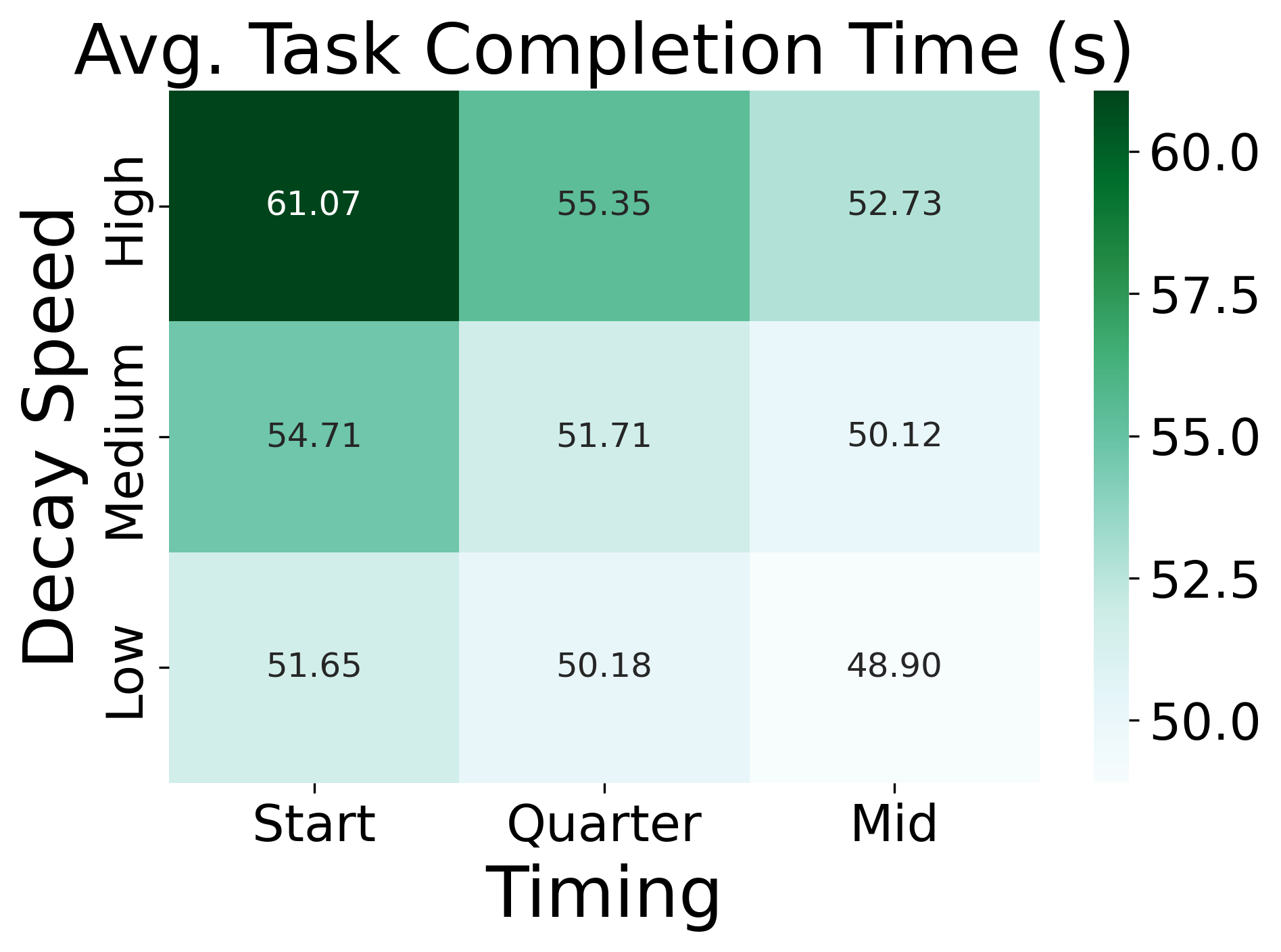}
    \caption{\textbf{Left:} TCT grouped by \textit{Timing} (x-axis) and \textit{Decay Speed} (color-coded), with baseline conditions \textit{Default} and \textit{Go-Go} included for comparison (shown in green and purple boxes, respectively).
    \textbf{Right:} Mean TCT across all participants, plotted by \textit{Timing} (x-axis) and \textit{Decay Speed} (y-axis).}
    \Description[Task completion time variation across experimental conditions and decay speeds.]{Task completion time variation across experimental conditions and decay speeds. Two complementary plots present TCT data: a box plot showing distribution of completion times and a heatmap displaying mean TCT values in seconds. The box plot reveals completion times ranging from approximately 40-120 seconds, with baseline conditions (Default and Go-Go) showing more consistent performance than experimental conditions. The heatmap demonstrates systematic patterns where high decay speed consistently requires longer completion times (52.73-61.07s) compared to low decay speed (48.90-51.65s). Task completion times generally decrease from start to mid timing conditions, with the fastest average completion (48.90s) occurring with low decay speed during mid timing, and the slowest (61.07s) with high decay speed during start timing.}
    \label{fig:TCT_comparison}
\end{figure}

\paragraph{\textbf{Task Completion Time}}

Significant main effects were found for both \textit{Timing} ($F_{2,168}=14.34$, p $<$ 0.001) and \textit{Decay Speed} ($F_{2,168}=16.69$, p $<$ 0.001) on TCT. No significant interaction effect was observed between these meta-parameters ($F_{4,168}=1.11$, p $=$0.35). Post-hoc tests revealed significant differences in TCT between "Start" and "Quarter" intervention timings, as well as between "Start" and "Mid" intervention timings (statistics are given in Table~\ref{tab:fatigue_stats_ART_ANOVA_posthoc}). Although no significant difference was found between "Quarter" and "Mid" timings, TCT generally increased as \textit{Timing} increased. Additionally, pairwise comparisons showed that TCT increased significantly with \textit{Decay Speed}, with the lowest levels observed in the "Low" condition and highest in the "High" condition. Figure~\ref{fig:TCT_comparison} presents both the distribution (left) and mean values (right) of TCT across all nine AlphaPIG conditions, as well as baseline conditions \textit{Default} and \textit{Go-Go} for comparison.

A significant effect of \textit{Interaction Technique} on TCT was found ($\chi^2_{10}$=54, p$<$0.001, N=242). Only one AlphaPIG condition demonstrated significantly higher TCT compared to \textit{Go-Go}: \textit{Mid-Low} (p $<$ 0.05) and two conditions exhibited significantly higher TCT than \textit{Default}: \textit{Start-High} (p $<$ 0.05) and \textit{Quarter-High} (p $<$ 0.05). A significant difference in TCT was found between \textit{Default} and \textit{Go-Go} (p $<$ 0.001), with mean completion times of 48 seconds and 57 seconds, respectively. Details on all \textit{Interaction Technique} post-hoc tests can be found in Table~2 in the Supplementary Material.

\begin{table*}[!h]
    \centering
    \resizebox{\textwidth}{!}{%
    \begin{minipage}{\textwidth}
    \centering
    \begin{tabular}{c|c|c|c|c|c|c|c|c}
    \textbf{DV} & \textbf{IV} & \textbf{Post-hoc} & \textbf{Mean 1} & \textbf{Mean 2} & \textbf{t-ratio} & \textbf{Adj. p-value} & \textbf{Sign.} & \textbf{Trend} \\ \hline \hline
	\multirow{6}{*}{Cum. Fatigue} & \multirow{3}{*}{Timing} & Start vs. Quarter & 10.57 & 11.97 & -4.726 & %4.82e-06 
    0.0000 & *** & < \\
	~ & ~ & Quarter vs. Mid & 11.97 & 13.62 & -5.783 & %6.98e-08 
    0.0000 & *** & < \\
	~ & ~ & Start vs. Mid & 10.57 & 13.62 & -10.509 & %1.07e-19 
    0.0000 & *** & < \\ \cline{2-9}
	~ & \multirow{3}{*}{Decay Speed} & Low vs. Medium & 13.13 & 11.47 & 5.407 & %4.33e-07 
    0.0000 & *** & > \\
	~ & ~ & Medium vs. High & 11.47 & 11.56 & 0.577 & 0.5646 & ~ & > \\
	~ & ~ & Low vs. High & 13.13 & 11.56 & 5.984 & %3.83e-08 
    0.0000 & *** & > \\ \hline

        \multirow{6}{*}{TCT} & \multirow{3}{*}{Timing} & Start vs. Quarter & 50.58 & 52.41 & 3.653 & %6.92e-04 
        0.0000 & *** & > \\
	~ & ~ & Quarter vs. Mid & 52.41 & 55.81 & 1.566 & 0.1192 & ~ & > \\
	~ & ~ & Start vs. Mid & 50.58 & 55.81 & 5.219 & %1.57e-06 
    0.0000 & *** & > \\ \cline{2-9}
	~ & \multirow{3}{*}{Decay Speed} & Low vs. Medium & 50.25 & 52.18 & -2.209 & 0.0285 & * & < \\
	~ & ~ & Medium vs. High & 52.18 & 56.38 & -3.519 & 0.0011 & ** & < \\
	~ & ~ & Low vs. High & 50.25 & 56.38 & -5.728 & %1.37e-07 
    0.0000 & *** & < \\ \hline
    
    \multirow{6}{*}{Q1} & \multirow{3}{*}{Timing} & Start vs. Quarter & 0.13 & 0.71 & -2.875 & 0.0091 & ** & < \\
	~ & ~ & Quarter vs. Mid & 0.71 & 1.00 & -1.218 & 0.2248 & ~ & < \\
	~ & ~ & Start vs. Mid & 0.13 & 1.00 & -4.094 & 0.0002 & *** & < \\ \cline{2-9}
	~ & \multirow{3}{*}{Decay Speed} & Low vs. Medium & 1.12 & 0.62 & 2.234 & 0.0268 & * & > \\
	~ & ~ & Medium vs. High & 0.62 & 0.11 & 2.716 & 0.0146 & * & > \\
	~ & ~ & Low vs. High & 1.12 & 0.11 & 4.951 & %5.36e-06 
    0.0000 & *** & > \\ \hline

	\multirow{3}{*}{Q2} & \multirow{3}{*}{Decay Speed} & Low vs. Medium & -0.95 & -0.55 & -2.841 & 0.0101 & * & < \\
	~ & ~ & Medium vs. High & -0.55 & -0.27 & -0.689 & 0.4921 & ~ & < \\
	~ & ~ & Low vs. High & -0.95 & -0.27 & -3.529 & 0.0016 & ** & < \\ \hline

    \multirow{6}{*}{Q3} & \multirow{3}{*}{Timing} & Start vs. Quarter & 0.47 & 0.95 & -3.253 & 0.0038 & ** & < \\
	~ & ~ & Quarter vs. Mid & 0.95 & 1.05 & -0.027 & 0.9785 & ~ & < \\
	~ & ~ & Start vs. Mid & 0.47 & 1.05 & -3.28 & 0.0038 & ** & < \\ \cline{2-9}
	~ & \multirow{3}{*}{Decay Speed} & Low vs. Medium & 1.21 & 0.85 & 2.860 & 0.0048 & ** & > \\
	~ & ~ & Medium vs. High & 0.85 & 0.41 & 3.111 & 0.0044 & ** & > \\
	~ & ~ & Low vs. High & 1.21 & 0.41 & 5.971 & %4.09e-08 
    0.0000 & *** & > \\ \hline

    \multirow{3}{*}{Q4} & \multirow{3}{*}{Timing} & Start vs. Quarter & 0.03 & -0.55 & 2.878 & 0.0090 & ** & > \\
	~ & ~ & Quarter vs. Mid & -0.55 & -0.82 & 2.026 & 0.0440 & * & > \\
	~ & ~ & Start vs. Mid & 0.03 & -0.82 & 4.905 & %6.58e-06
    0.0000 & *** & > \\ \hline
    \end{tabular}
    \caption{Post-hoc comparisons of \textit{Timing} and \textit{Decay Speed} effects on cumulative fatigue, task completion time (TCT), and body ownership/agency questionnaire responses (Q1-Q4). Mean values are provided for each meta-parameter level. P-values are adjusted using the Holm-Bonferroni method. Significance levels: ***: $p < 0.001$, **: $p < 0.01$, *: $p < 0.05$. In the Trend column, "$>$" indicates higher values for the left meta-parameter level, while "$<$" indicates higher values for the right meta-parameter level based on paired rank comparisons.}
    \label{tab:fatigue_stats_ART_ANOVA_posthoc}
    \end{minipage}
    }
\end{table*}

\paragraph{\textbf{Body Ownership and Agency}}
% We found consistent, significant effects of increasing the intervention time from "Start" (intervention at the beginning of an episode) to "Quarter" (intervention at 25\% of the reference fatigue level) on Q1, Q3 and Q4, suggesting that delaying the activation of the virtual offset between 0\% and 25\% has a positive effect on perceived body ownership and agency. We could not identify a consistent significant difference between ownership and agency perceived for "Quarter" and "Mid" intervention times. For the decay rate, we observed significant differences in Q1 and Q3 between all three parameter levels. However, no significant differences were observed in Q4, and only between "Low" and "Medium" and between "Low" and "High" in Q2. This means that we could not find an AlphaPIG condition with an improved perception of body ownership or agency compared to each of the two baselines, and only one AlphaPIG condition (\textit{Mid-Medium}) with some minor evidence for a degraded ownership perception.

\begin{figure}
    \centering
    \includegraphics[width=0.98\linewidth]{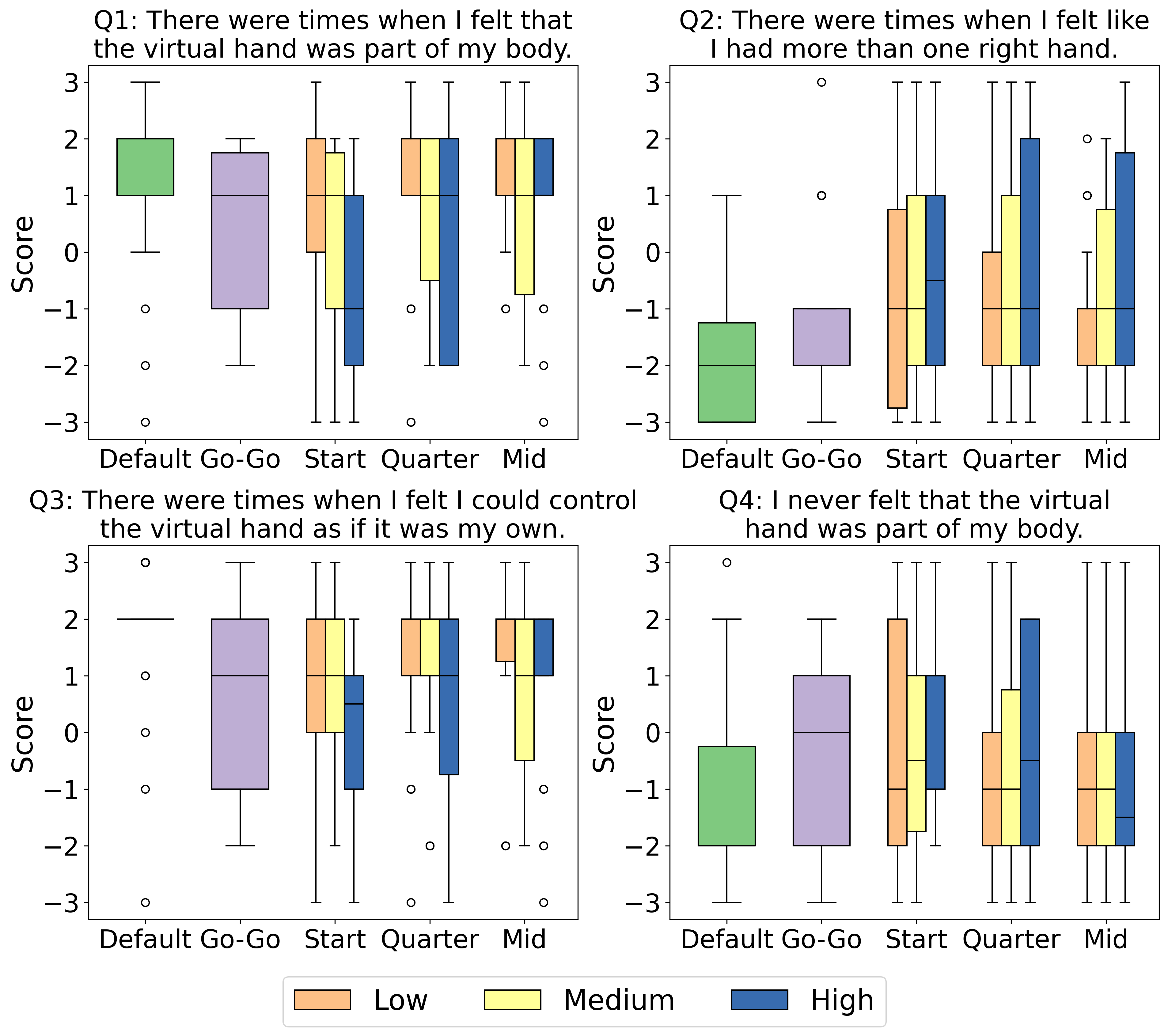}
    \caption{Questionnaire responses (Q1-Q4) grouped by \textit{Timing} (x-axis) and \textit{Decay Speed} (color-coded), with baseline conditions \textit{Default} and \textit{Go-Go} included for comparison (shown in green and purple boxes, respectively).}
    \Description[Analysis of user perception of virtual hand embodiment across four survey questions.]{Analysis of user perception of virtual hand embodiment across four survey questions. The figure presents box plots for four questions assessing different aspects of virtual hand perception, with scores ranging from -3 to +3. Q1 (body ownership) shows generally positive scores across conditions, with Default showing highest median scores. Q2 (multiple hand sensation) reveals negative scores for baseline conditions but varied responses in experimental conditions, particularly higher scores with high decay speed. Q3 (control perception) demonstrates consistently positive median scores across conditions, with experimental conditions showing wider response ranges. Q4 (body ownership negation) displays inverse patterns to Q1, validating response consistency. Experimental conditions (Start, Quarter, Mid) generally show greater response variability compared to baseline conditions (Default, Go-Go), with decay speed affecting response patterns differently across questions.}
    \label{fig:Questionnaire_comparison}
\end{figure}

\begin{figure}
    \centering
    \includegraphics[width=0.98\linewidth]{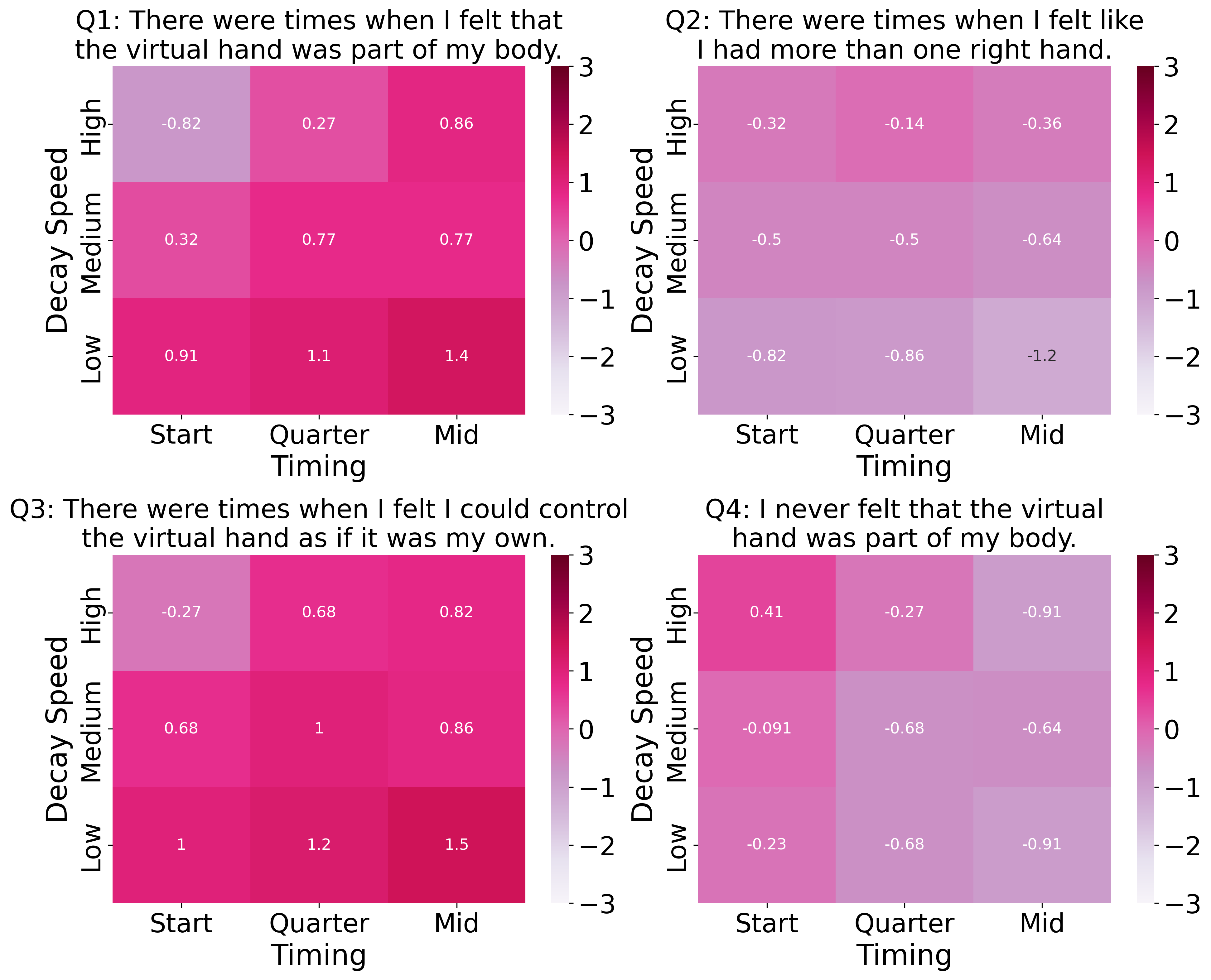}
    \caption{Mean questionnaire responses (Q1-Q4) across all participants, plotted by \textit{Timing} (x-axis) and \textit{Decay Speed} (y-axis).}
    \Description[Heatmap visualization of average virtual hand perception scores across experimental conditions.]{Heatmap visualization of average virtual hand perception scores across experimental conditions. Four heatmaps display mean responses (-3 to +3) for different aspects of virtual hand experience. Q1 (body ownership) shows highest positive scores with low decay speed (0.91-1.4), particularly during mid timing. Q2 (multiple hand sensation) reveals consistently negative responses across all conditions, with strongest negative perception (-1.2) at low decay speed and mid timing. Q3 (control perception) demonstrates increasingly positive responses toward mid timing, peaking at 1.5 with low decay speed. Q4 (body ownership negation) shows predominantly negative responses, validating Q1 findings, with strongest disagreement (-0.93) during mid timing at low decay speed. Overall patterns indicate low decay speed consistently produces stronger responses across all questions.}
    \label{fig:Questionnaire_comparison_heatmaps}
\end{figure}

Significant main effects were found for \textit{Timing} on Q1 ($F_{2,168}=8.84$, p $<$ 0.001), Q3 ($F_{2,168}=7.12$, p $<$ 0.01), and Q4 ($F_{2,168}=12.15$, p $<$ 0.001). Although no significant difference was found between "Quarter" and "Mid" timings for Q1 and Q3 in post-hoc tests, perceived body ownership generally increased as \textit{Timing} increased. Additionally, significant main effects of \textit{Decay Speed} were found for Q1 ($F_{2,168} =12.29$, p $<$ 0.001), Q2 ($F_{2,168}=7.00$, p $<$ 0.01) and Q3 ($F_{2,168}=17.84$, p $<$ 0.001). While post-hoc comparisons show no significant difference between "Medium" and "High" decay speeds for Q2, perceived body ownership generally decreased as \textit{Decay Speed} increased. An interaction effect between \textit{Timing} and \textit{Decay Speed} was only found for Q1 ($F_{4,168}=3.11$, p $<$ 0.05). Post-hoc comparisons of individual parameter effects are presented in Table~\ref{tab:fatigue_stats_ART_ANOVA_posthoc}. Figure~\ref{fig:Questionnaire_comparison} and~\ref{fig:Questionnaire_comparison_heatmaps} presents the distributions and mean values of post-study questionnaires across all nine AlphaPIG conditions, as well as baseline conditions \textit{Default} and \textit{Go-Go} for comparison.

A significant effect of \textit{Interaction Technique} was found on scores across all four questionnaire items (Q1: $\chi^2_{10}$=$52$, Q2: $\chi^2_{10}$=$37$, Q3: $\chi^2_{10}$=$38$, Q4: $\chi^2_{10}$=$33$; all with p$<$0.001 and N=242). Pairwise post-hoc comparisons revealed no significant differences among the 11 conditions, with only Q2 showing significantly higher scores for \textit{Mid-Medium} compared to \textit{Default} (p $<$ 0.05; see Tables 3-6 in the Supplementary Material). These findings demonstrate that through systematic adjustment of two meta-parameters—\textit{Timing} and \textit{Decay Speed}—AlphaPIG effectively preserves comparable levels of body ownership and agency to non-adaptive \textit{Go-Go}.

\subsection{Perception of Manipulation}
\label{sec:quantitativeresults}
Our post-study questionnaire included an open-ended item exploring participants' perception of in-study manipulation (Q5: "Do you notice any changes during the interaction time?"). We used quantitative content analysis~\cite{petrovskaya2022prevalence,vears2022inductive} to count the occurrence of "Lack of awareness of the interaction technique" responses marked as "no", left blank, or indicating passive acceptance of the virtual hand. Among all responses, 38 out of 240 responses indicated no perceived changes in the interaction technique during the trial.

Twelve participants reported no notable difference in the \textit{Default} condition. They described the \textit{Default} condition as offering \textit{“very immediate interaction”} (P1) and being \textit{“by far the best run”} (P20). However, two participants (P5, P10) noted that they had to exert more effort for buttons positioned farther from their chest.

Seven participants reported negative perceptions of manipulation in the \textit{Mid-Low} condition. While some participants characterized it as \textit{“very fast”} (P1), \textit{“syncs well with my real hands”} (P13, P16, P20), and \textit{“easy to control”} (P9), those who noticed the manipulation during the latter half of the interaction expressed mixed opinions regarding hand control (P4, P6, P8, P11, P14, P15, P20). For example, P8 noted \textit{“I can feel the hand extension, but [it is] not too hard to control”}, whereas P14 found it \textit{“very hard to control during the second half of the process”}.

Four participants reported negative perceptions of \textit{Go-Go}, while four others reported negative responses to the \textit{Quarter-Low} condition. For Go-Go, seven participants (P3, P4, P9, P13, P14, P15, P17) found the manipulation noticeable but \textit{“easy to use”}. Their comments included \textit{“I noticed a small distance between my actual hand and the virtual hand from the very beginning”} (P12), \textit{“I could treat it as an extension of my hand”} (P3), and \textit{“the virtual hand feels like a controller”} (P17). Regarding \textit{Quarter-Low}, four participants (P1, P3, P11, P17) described it as \textit{“easy”}, noting that \textit{“Shorter hand extension was more manageable to control”} (P3) and \textit{"Towards the end there was a little bit of offset but overall it felt easy"} (P17).

The \textit{Quarter-High}, \textit{Start-Medium}, \textit{Mid-Medium}, and \textit{Start-Low} conditions each received 1-3 negative responses, with participants expressing mixed opinions. Positive feedback highlighted improved control over time: \textit{“target pointing became easier towards the end [in the \textit{Quarter-High} condition]”} (P6), and \textit{“I feel the virtual hand is less like a part of my body at the beginning. Later, I had more control, and [the virtual hand] feels more [like] a part of my body [in the \textit{Start-Medium} condition]”} (P13). Some participants noted natural movement, with one stating \textit{“the movement of the virtual hand [in the \textit{Start-Low} condition] is like my [real] hand”} (P20). However, others reported control issues: \textit{“Sometimes I could control the virtual hand well [in the \textit{Start-Low} condition], but sometimes it was out of my control”} (P18) and \textit{“The hand [in the \textit{Quarter-High} condition] became a bit offset, making it feel less immersive”} (P17).

%% file: sections/5_discussion.tex
\section{Discussion}
Our implementation demonstrates AlphaPIG's effectiveness when integrated with an established hand redirection technique for mid-air selection tasks. Both quantitative and qualitative findings confirm that AlphaPIG successfully transforms the existing redirection offset-ownership trade-off into a fatigue-ownership trade-off (\textbf{H1}, \textbf{H2}, \textbf{H3}, \textbf{H4}) in fatigue-aware interaction design. Moreover, AlphaPIG achieves an optimal balance, maintaining comparable levels of ownership while significantly reducing fatigue.

\subsection{The Physical Fatigue and Body Ownership Trade-off}
Our case study from Section~\ref{sec:casestudy} investigates the parameter tuning of AlphaPIG to establish foundational insights into how intervention timing and intervention decay speed affect the trade-off between shoulder fatigue and body ownership. These findings provide essential design considerations for implementing novel interaction techniques in future XR applications.

As mentioned in Sec~\ref{sec:redirection-technique}, previous research~\cite{wentzel2020improving} has identified a trade-off between increased virtual hand offset and reduced perceived body ownership when using hand redirection to minimize arm movement. In our implementation of AlphaPIG-assisted Go-Go, we dynamically adjust the distance threshold $\theta$ based on real-time fatigue predictions, providing a practical interpretation of redirection intensity. This approach enables AlphaPIG to explore manipulation parameters within the context of the physical fatigue and body ownership trade-off.

\begin{figure}[!h]
    \centering
    \includegraphics[width=0.98\linewidth]{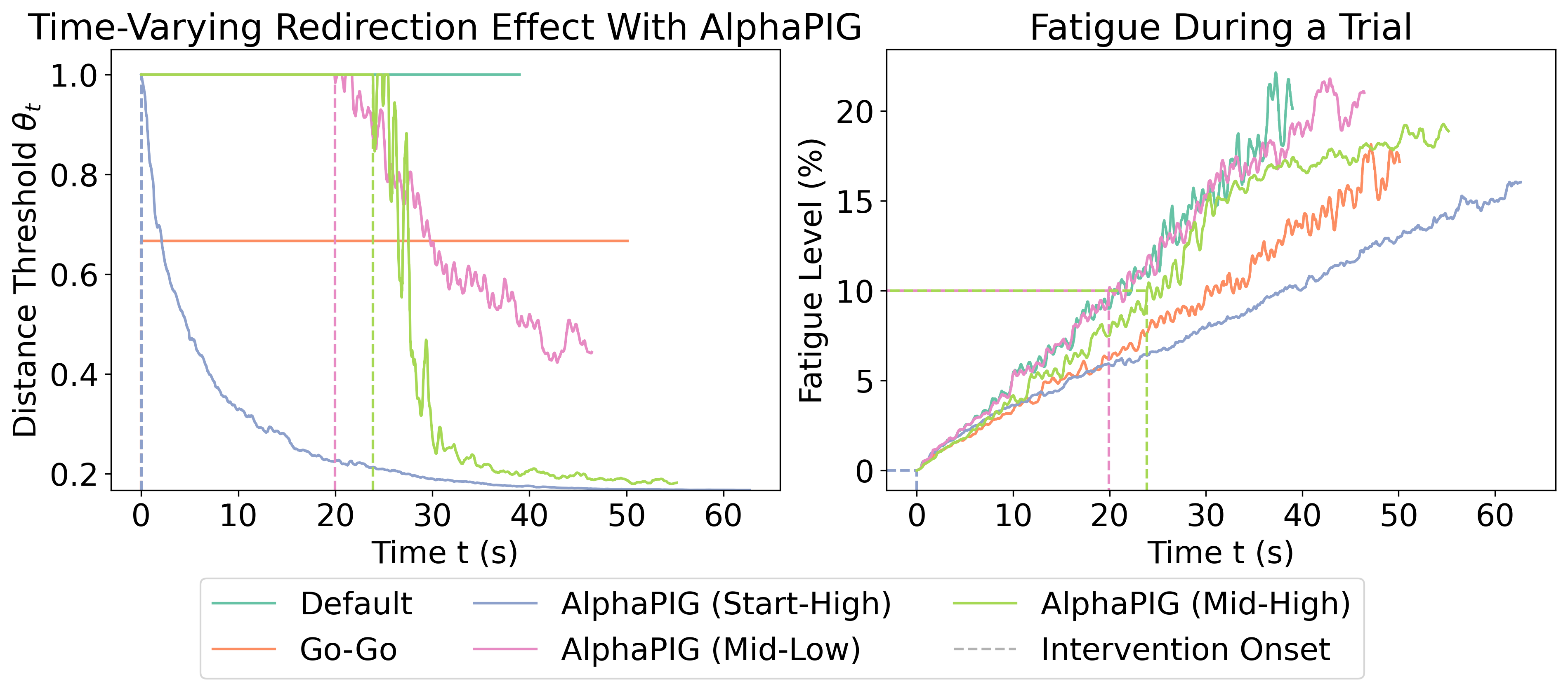}
    \caption{With AlphaPIG, different types of distance threshold (left) and fatigue (right) trajectories can be realised. For example, applying the redirection effect from the beginning (\textit{Start-High}, blue lines) results in a smaller final fatigue than delaying it until a certain fatigue threshold has been reached (\textit{Mid-Low}, magenta line, with intervention onset at F=10 in this case).
    Exemplary trajectories were taken from a single participant (P14).}
    \Description[Comparison of distance threshold and fatigue trajectories under different AlphaPIG configurations.]{Comparison of distance threshold and fatigue trajectories under different AlphaPIG configurations. This dual-plot figure presents time-series data from participant P14 over 60 seconds. The left plot shows distance threshold (theta) changes, where Start-High demonstrates early rapid decay to 0.2, while Mid-Low and Mid-High conditions maintain higher thresholds until their respective intervention points. The right plot displays corresponding fatigue development, revealing that early intervention (Start-High) results in gradual fatigue accumulation reaching approximately 15\%, compared to steeper increases in delayed intervention conditions. Baseline conditions (Default and Go-Go) provide reference trajectories, with Go-Go maintaining a constant threshold of 0.7 and Default showing rapid early adaptation.}
    \label{fig:intervention_effect_phaseI_example}
\end{figure}

Our analysis revealed a positive correlation between intervention timing and accumulated fatigue during trials. As shown in the left of Figure~\ref{fig:intervention_effect_phaseI_example}, the later timing in the \textit{Mid-High} condition enabled the distance threshold $\theta$ to be decreased to the same level as the \textit{Start-High} condition, but achieving this within a short time frame. Consequently, the right side of Figure~\ref{fig:intervention_effect_phaseI_example} reveals that the growth rate of cumulative fatigue in \textit{Mid-High} parallels that of \textit{Start-High} during the second half of interaction. However, due to time constraints, \textit{Mid-High} concluded with higher overall fatigue levels than \textit{Start-High}. These findings suggest that earlier interventions maximize effectiveness in fatigue reduction. Conversely, delayed interventions prove less effective, as accumulated fatigue becomes more difficult to mitigate within the remaining interaction time.

Analysis revealed a negative correlation between \textit{Decay Speed} and cumulative shoulder fatigue. The left side of Figure~\ref{fig:intervention_effect_phaseI_example} shows that the distance threshold $\theta$ in the \textit{Mid-High} condition remains lower than in the \textit{Mid-Low} condition. Consequently, the higher decay speed in \textit{Mid-High} leads to slower fatigue accumulation, while the lower decay speed in \textit{Mid-Low} results in more rapid fatigue build-up, as illustrated in Figure~\ref{fig:intervention_effect_phaseI_example}. Our case study provides strong evidence that fatigue reduction can be achieved through two principal mechanisms: earlier intervention timing and increased physical-virtual hand offset.

However, these fatigue reduction strategies may increase TCT. Figure~\ref{fig:intervention_effect_phaseI_example} shows that \textit{Mid-High} exhibits longer TCT than \textit{Mid-Low} while remaining shorter than \textit{Start-High}. These observations are supported by our quantitative analysis in Section~\ref{sec:quantitativeresults}, which reveals significant TCT increases with higher decay rates and earlier interventions (e.g., 50 seconds for \textit{Quarter-Low} versus 55 seconds for \textit{Quarter-High}). The increased hand redirection offset appears to introduce complexities in controlling the extended virtual hand, requiring additional time for participants to adapt their movements. Participant feedback highlighted the occasional control difficulties and reduced immersion during hand redirection, aligning with questionnaire responses. We observed a slight decline in perceived body ownership and agency with decreasing $\theta$, suggesting that rapid changes in interaction technique may compromise user engagement and immersion. These findings point to a key design principle: gradually introducing interventions after sustained interaction may enhance users' adaptation capabilities while preserving their sense of embodiment.

In addition, we have shown that both \textit{Decay Speed} and \textit{Timing} serve as effective parameters for balancing these opposing design goals. Two conditions (\textit{Start-Medium} and \textit{Quarter-Medium}) achieved significantly reduced fatigue compared to the default direct interaction without compromising perceived ownership and agency. Notably, the \textbf{Start-Medium} condition demonstrated significantly lower fatigue than the non-adaptive Go-Go technique without fatigue-based intervention. This demonstrates that AlphaPIG can help optimize the Go-Go technique by effectively mitigating cumulative fatigue during mid-air pointing tasks, utilizing meta-parameters derived from our systematic trade-off exploration. Furthermore, AlphaPIG facilitates the iterative refinement of fatigue-aware XR applications: designers can explore additional parameter combinations around optimal conditions (e.g., intervention onsets between 0\% and 25\%, decay rates between 0.25 and 0.45).

\subsection{Guidelines for Using AlphaPIG}
\label{sec:guideline}
AlphaPIG lowers the barrier to creating fatigue-aware XR applications by providing an easy way to adapt and improve hand-based interaction techniques for future XR products. Here are three essential considerations for applying AlphaPIG in a wider range of applications effectively:

\paragraph{Interaction Task}
Task characteristics—specifically cognitive and physical sparsity— influence the selection of parameters $T_{f}$ and $DR_{\alpha}$ in the AlphaPIG system. In our AlphaPIG-assisted Go-Go implementation, the mid-air pointing task exhibited low sparsity in body movement and mental demands, characterized by participants sequentially pressing buttons without temporal delays. Our study results found that a small $T_{f}$ and a large $DR_{\alpha}$ might cause a loss of interaction control, despite strategically timing display updates to align with the human blink rate (every 3.33 seconds) to minimize perceptual disruption. We suggest applying an intervention timing between $25-50\%$ of $F_{max}$ and maintaining a low $DR_{\alpha}$ between $0.1-0.25$. These parameters are critical for preserving interaction continuity in dense interaction scenarios. Furthermore, integrating attention management tools~\cite{lingler2024supporting} would help optimize the intervention timing and thus further enhance user experience. 

In contrast, in tasks characterized by high cognitive and physical sparsity, such as doing immersive analytics with large datasets, participants may intermittently pause their physical movements to support cognitive processing and analytical thinking. This complexity necessitates a bidirectional approach to manipulation parameters: decreasing $\theta_t$ during active task engagement and incrementally increasing it during a temporal cognitive pause. A small $T_{f}$ ($\approx0-25\% \cdot F_{max}$) combined with a high $DR_{\alpha}$ between 0.25 and 0.45 enables a responsive adaptation to fatigue variation without compromising task continuity. 

\paragraph{Interaction Technique}
Interaction techniques implemented with AlphaPIG require careful consideration to identify parameters suitable for fatigue-aware interventions. As outlined in Section~\ref{sec:novelmanipulation}, hand-based interaction techniques can enhance ergonomics across three dimensions, each supported by AlphaPIG's design. While our case study focused on the Go-Go activation threshold, AlphaPIG's application extends to other interaction techniques. For instance, in target placement optimization based on arm movement, as explored in~\cite{evangelista2021xrgonomics}, AlphaPIG can determine optimal intervention timing for target relocation and guide incremental position updates.

\begin{figure}[!h]
    \centering
    \includegraphics[width=\linewidth]{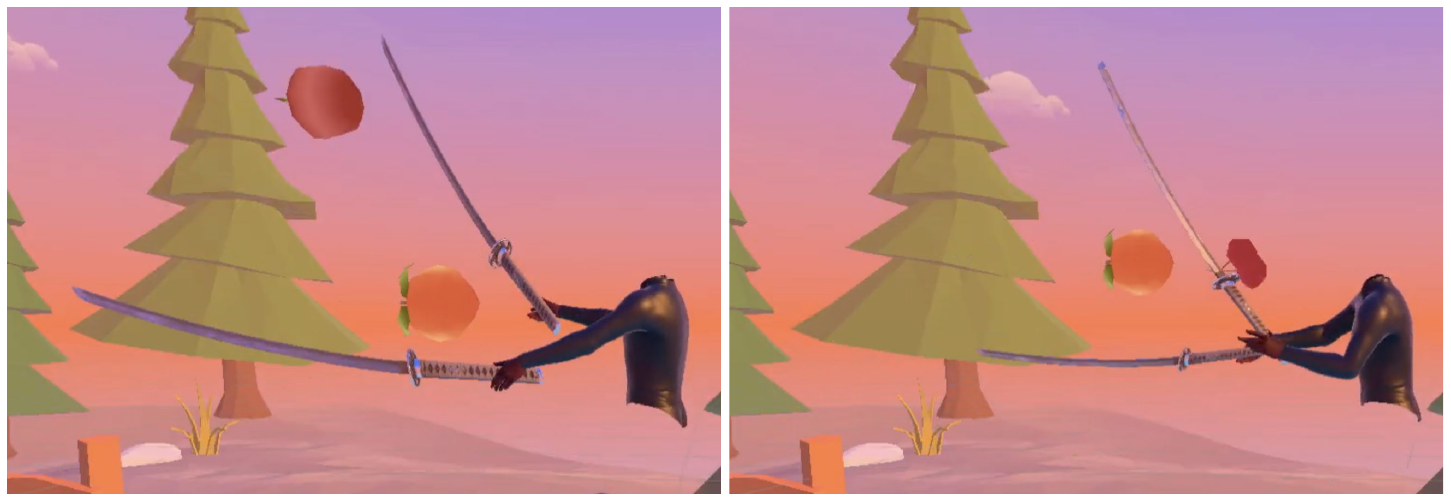}
    \caption{Demonstration of AlphaPIG's fatigue modulation in an exertion game. Left: Default blade length enables easy fruit slicing. Right: Reduced blade size increases task difficulty when the fatigue falls below the threshold.}
    \Description[Visual comparison of adaptive difficulty in a virtual reality fruit-slicing game.]{Visual comparison of adaptive difficulty in a virtual reality fruit-slicing game. Side-by-side images demonstrate AlphaPIG's fatigue-based blade length modulation. The left image shows a player wielding a full-length sword for optimal interaction range, while the right image illustrates the automatic blade length reduction mechanism triggered by detected fatigue, requiring more precise movements to accomplish the same fruit-slicing task. Both scenes share identical environmental elements including stylized pine trees and a sunset-colored sky, emphasizing that only the interaction mechanic changes.}
    \label{fig:fruitninja}
\end{figure}

Beyond improving interaction ergonomics through fatigue reduction, AlphaPIG can actively modulate fatigue levels to enhance engagement in exertion games. For instance, when applied to "Fruit Ninja"\footnote{The example implementation was based on an open-source VR game developed by KutyVr (\url{https://github.com/KutyVr/Fruit-Ninja-VR}).}, AlphaPIG can dynamically adjust blade size ($\theta$) based on player fatigue levels. During periods of low exertion (i.e., fatigue below threshold), AlphaPIG reduces blade size to increase selection difficulty (see Figure~\ref{fig:fruitninja}).

\paragraph{User Experience Trade-off}
Implementation of AlphaPIG across different interaction techniques requires careful consideration of trade-offs between physical fatigue and other user experience factors. Our case study with the Go-Go technique examined the balance between shoulder fatigue and body ownership. Similarly, when applying AlphaPIG to target relocation, designers must consider the trade-off between physical fatigue and task performance, as dynamically updating the target position may compromise task accuracy.

\subsection{Limitations and Future Work}
While our study provides valuable insights for fatigue-aware interaction design, several limitations suggest directions for future research.

First, participants reported control difficulties in our AlphaPIG-assisted Go-Go implementation when interaction techniques underwent substantial changes within brief periods. Despite the implementation of smoothing factor $\beta$, rapid technique modifications still contributed to increased task completion times. Additionally, our study was limited to experienced VR users due to their familiarity with interaction control and body ownership concepts. Future research should explore enhanced communication methods for transitions between default and manipulated interaction techniques, such as visual cues or tactile feedback while evaluating their effectiveness across both novice and experienced users. Furthermore, examining various smoothing factors $\beta$ could optimize transitions between isomorphic and anisomorphic interaction, potentially minimizing disruptions.

Second, our AlphaPIG implementation relies on the NICER shoulder fatigue model, inherently limiting its application to upper body movements. Future iterations should incorporate additional fatigue measurements beyond the upper body to enhance generalizability.

Third, our study was limited to single-arm interactions with right-handed participants to directly assess body ownership and maintain study consistency. Future research should examine bimanual tasks to investigate intervention effects on two-handed coordination. Additionally, while the current study visualized only the right hand, future work should explore full-body virtual avatars and employ comprehensive embodiment surveys~\cite{roth2020construction}.

Fourth, the current AlphaPIG implementation uses an exponential function to intensify interventions when real-time fatigue exceeds predetermined thresholds. Future research should investigate alternative continuous functions, such as quadratic or sigmoid models, to optimize intervention strategies based on user perception of manipulation.

%% file: sections/6_conclusion.tex
\section{Conclusion}
This paper introduces \textit{AlphaPIG}, a meta-technique designed to \textbf{P}rolong \textbf{I}nteractive \textbf{G}estures through real-time fatigue management during mid-air interaction, leveraging state-of-the-art shoulder fatigue modelling. Through two meta-parameters—fatigue threshold and decay rate—AlphaPIG modulates the timing and intensity of ergonomic interaction techniques, enabling XR designers and researchers to explore trade-offs between fatigue reduction and user experience, thus simplifying the implementation of novel ergonomic interactions in fatigue-aware XR design. Using the established Go-Go interaction technique as a case study, we demonstrate AlphaPIG's ability to facilitate systematic exploration of intervention effects in hand redirection, particularly the balance between muscle fatigue and body ownership. Based on these findings, we provide guidelines for implementing AlphaPIG across diverse interaction techniques and tasks.

%% file: sections/7_openscience.tex
\section*{Open Science}
In support of open science principles, we are committed to making AlphaPIG and the AlphaPIG API readily accessible to enhance transparency, reproducibility, and collaboration. AlphaPIG utilizes the openly available NICER shoulder fatigue model~\cite{li2024nicer} to optimize user comfort and interaction efficiency. We provide comprehensive documentation and release the meta-technique and API under an open-access license via GitHub, enabling researchers to reproduce results, contribute improvements, and explore new applications. The AlphaPIG Unity plugin and the anonymized study data are available at: \url{https://github.com/ylii0411/AlphaPIG-CHI25}. %, along with the anonymized data from the case study.